# Quantum Optical Aspects of High-Harmonic Generation.

**Sándor Varró**

ELI-ERIC, ALPS (Attosecond Light Pulse Source) Research Institute,
ELI-HU; 6728 Szeged, Wolfgang Sandner utca 3, Hungary; E-mail: varro.sandor@eli-alps.hu

**Abstract:** The interaction of electrons with strong laser fields is usually treated with semiclassical theory, where the laser is represented by an external field. There are analytic solutions for the free electron wave functions, which incorporate the interaction with the laser field exactly, but the joint effect of the atomic binding potential presents an obstacle for the analysis. Moreover, the radiation is a dynamical system, the number of photons changes during the interactions. Thus, it is legitimate to ask how can one treat the high order processes nonperturbatively, in such a way that the electron-atom interaction and the quantized nature of radiation be simultaneously taken into account? An analytic method is proposed to answer this question in the framework of nonrelativistic quantum electrodynamics. As an application, a quantum optical generalization of the strong-field Kramers-Heisenberg formula is derived for describing high-harmonic generation. Our formalism is suitable to analyse, among various quantal effects, the possible role of arbitrary photon statistics of the incoming field. The present paper is dedicated to the memory of Prof. Dr. Fritz Ehlotzky, who had significantly contributed to the theory of strong-field phenomena over many decades.

## 1. Introduction

The process of high-order harmonic generation has been a subject of extensive theoretical investigation since the 60ies of the last century, almost immediately after the laser was invented. The main challenge has been to treat the interaction with a strong laser radiation nonpertubatively. This is possible, on the basis of the famous Gordon-Volkov states [1,2], which are exact solutions of the Klein-Gordon or Dirac equation of a charged particle in a classical electromagnetic plane wave. The high-harmonic production in nonlinear Compton scattering has been studied by several authors [3–6]. In [6] the fully quantized description has been given, thanks to the analytic solution of the Dirac equation in a quantized electromagnetic plane wave [7]. In [8] the nonrelativistic version of [7] has been applied to the process of induced multiphoton Bremsstrahlung. The nonrelativistic (semiclassical) Gordon-Volkov states have been used in the Keldysh-Faisal-Reiss model of ionization in strong laser fields [9–18], as well as in the theory of induced Bremsstrahlung in laser-assisted electron scattering [19–27]. Like the first experimental demonstration of above-threshold ionization [14], the observation of high-order harmonic generation [28–30] gave again a fruitful impetus to the development of theoretical models [31–38]. In the meantime a proposal appeared on the generation of attosecond light pulses from the high harmonics [39], which relied on the much debated assumption that the harmonic components are phase-locked. After the first experimental indication [40], attosecond pulse trains [41,42], and isolated pulses have been observed, as well [43]. In the last two decades a new sub-discipline, attosecond physics [43–48] was born, whose various applications are also covered by international projects, like the ELI-ALPS project [49,50].

The common feature of most theoretical descriptions of above-threshold ionization (ATI), high-harmonic generation (HHG), has been to take into account the effect of the high-intensity laser field nonperturbatively, in terms of the Gordon-Volkov states, which are *dressed free electron states*, and the other effects are treated as perturbations. This is just the "Keldysh philosophy", being the essence of the "strong-field-approximation" (SFA). At this place we would like to highlight the early paper by Ehlotzky [33], which seems to be the very first one in discussing HHG, on the basis of S-matrix theory in the Keldish spirit. The SFA has been very successful in interpreting many features of strong-field phenomena during decades [37,51–54]. This success stems from the fact that the Volkov states are exact quasi-classical states. In other words, the "WKB approximation" is not an approximation for Volkov states, so in such a fortunate situation quantum mechanics can intuitively, and correctly be formulated in classical terms. This is a special case demonstrating Van Vleck' general theorem [55], according to which for quadratic Hamiltonians the quasi-classical description is exact. As a result, the Feynman path-integral is stationary at the classical action [56], which is just contained in the Volkov propagator.

Besides the extensive numerical work, a considerable theoretical research has been carried out for an *analytic* formulation, which would go beyond the strong field approximation. In such a theory the nonperturbative description of the laser-electron interaction would be kept, of course, but, at the same time, the simultaneous effect of the nucleus would also be accurately incorporated. For instance, the "Coulomb tail" in the laser-dressed scattering states or final states of ionization has already long been studied, in terms of the "Coulomb-Volkov waves" [57–61], which are space-translated continuum states. Since the "p.A"-term contains a gradient, in this descriptions, too, the space-translated oscillating potential quite naturally appears [12,62–69]. Using an analogous procedure to that of Kramers [62] and Henneberger [63], we have worked out a general formalism [36] for treating high-intensity multiphoton processes, which goes well beyond the strong field approximation. As an application of this formalism, we have treated high harmonic generation on atoms, and derived the "multiphoton Kramers-Heisenberg formula" [36], in the spirit of S-matrix theory of light scattering. This is a semiclassical formula (the laser light has been represented by an external field), but the method can be generalized for the case of quantized fields. Our main purposes in the present paper is to derive a quantized version of this formula, which will be done in the frame of nonrelativistic quantum electrodynamics.

Now let us deal with some conceptual points connected to the quantized description of HHG (or, in general, high-order processes), and briefly overview some earlier and recent works. It is well known that in the semiclassical interpretation of "multiphoton processes" the word "photon" appears at the stage, when one encounters nonlinear transitions matrix elements (higher powers of the oscillating external field) having electron's phase factors of the type $\exp[(2\pi i/h)(E_f - E_i)\cdot t]\times\exp(-2\pi i n\nu\cdot t)$. By an elementary operation, this expression can be brought to an equivalent form, $\exp\{(2\pi i/h)[E_f - (E_i + nh\nu)]\cdot t\}$. At resonance the initial and final energies, $E_i$ and $E_f$, respectively, are connected by the formula $E_f = E_i + nh\nu$, where $h$ is the Planck constant, and then we say "the electron absorbs *n* photons". However, in this description the Planck constant is not a "property" of the radiation (consisting of photons of energy $h\nu$), but it belongs to the de Broglie–Schrödinger wave of the electron. The other extreme approach would be to take an external current, and see to its quantized radiation field. Motivated partly by the work of Bloch and Nordsieck [70], Glauber [71] and Schwinger [72] have shown that an external (classical) current distribution generates coherent states from the vacuum state. In accord with this theorem, an external current of electrons oscillating under the action of a strong laser field generates high-harmonic radiation being in a multimode (product) coherent state [73,74]. Each components have a Poisson photon number distribution, and their phases are locked, so, in principle, attosecond pulses [39] could be formed by interference. In contrast to this, in the semiclassical model of the *quantum* dipole radiators of HHG the phase-locking does not come out automatically. The phase-differences of the consecutive harmonics are stable only in a certain range of the spectrum [75], as is also the case for the ATI electron de Broglie waves [76]. If one goes over to the quantized description of the radiation, then an initially localized free electron and the field find themselves in an entangled state [77]. Interestingly, the resulting (quite non-classical) photon number distribution depends on the position of the detection of the electron [77]. Moreover, there may be an accumulation of the entanglement entropy by the end of the interaction, depending on the smoothness of the switching. In case of a sine squared switching function (which is often used in numerical simulations) the system is almost recovering to a pure state, i.e., it makes an *almost reversible* cycle [78]. Recent studies [79–83] have led to a conclusion that in the process of high-harmonic generation the Mandel Q parameter becomes only slightly different from 0, if the incoming field is in a coherent state. In [80] and [82] only super-poissonian (>0) values has been found for the fundamental component, in [83], depending on the order of the harmonic, also sub-poissonian (<0) values came out for the photon number distributions, which is a signature of non-classical fields. It is interesting to note that if one takes symmetrically entangled multimode coherent states of (say) the plateau harmonics, whose coherent state parameters have random phases, still one has an attosecond locking in the expectation value of the electric field strength [81]. So, one easily sees that the introduction of the quantized nature of radiation, immediately results in uncountable many phenomena which does not exist in the classical domain. Let us just mention that the classical versus quantum aspects of free-electron lasers (see references at the end of Appendix B) has also been the subject of extensive discussion from the very beginning.

Since the radiation is a dynamical system, the number of photons changes during the interactions, e.g., the incoming fundamental radiation loses photons. This *depletion* is used out in the development of the principle and design of the quantum spectrometer [84–87]. The method is based on measuring the difference signal between the incoming and outgoing fundamental radiation, which previously induced the emission of high harmonics. In this way the high-harmonic spectrum can be recovered quite accurately from measuring the number of “missing photons”. Let us mention here that the effect of depletion has already been described in [6], in the context of high-harmonic generation in a Compton process. A surprising conclusion of that study was that the depletion can cause even kinematic effects (like additional frequency shift).

Recently several theoretical works have appeared, which discuss the question of how the quantum nature of light manifest itself in the high-order multiphoton processes [80,82,83,88–91], and offer various solutions. In a detailed investigation of the subject of high-harmonic generation [83] the authors explicitely consider the effect of the whole quantized radiation field. In [83] a coherent state part has been separated for the laser mode, and in this way, on one hand, it was possible to incorporate the exact semiclassical dynamics (which can be handled numerically), and, on the other hand, the “rest” genuine quantal features could also be accurately described. In our treatment below, we also consider the whole quantized radiation field, however we do not use any *Ansatz* from the outset (like coherent states for the incoming field). Of course, on physical grounds, it is justified to pick up the strong modes, however these need not be in ideal coherent states. In our formalism the initial states of the whole quantized field is *arbitraty*, thus, we can consider e.g., squeezed coherent or thermal incoming fields. After all, the strong fields used in the experiments are usually generated by parametric amplification, e.g., OPCPA [49,50], so the strong field impinging on the target is surely not in an ideal coherent state. The present work offers an analytic method which goes well beyond the strong field approximation, even in the case when this would be applied with quantized fields. Our formalism, which is based on the elimination technique used in [6,8], nonperturbatively describes the joint interaction of the electron both with quantized radiation and with the atomic potential. With the applied method we cannot handle the mode-mode coupling of *several strong* modes *exactly*, so this sets the basic limit of the physical applicability of some of our results. As an application, the quantum version of the semi-classical “strong-field Kramers-Heisenberg formula” of [36] shall be derived, which seems to be a suitable tool to analyse, among various quantal effects, the role of arbitrary photon statistics of the incoming high-intensity radiation. We would like to highlight the conceptual differences and similarities of the external field approximation and the quantum optical description, on the basis of the proposed formulation, and we shall not make quantitative comparisons with particular experiments. The emphasis will be put on the various theoretical approaches, and some novel phenomena in strong-field physics, whose interpretation needs quantum optics.

In Section 2 we set up a first-principle theoretical model based on nonrelativistic quantum electrodynamics. We shall eliminate the minimal coupling interaction terms by using the squeezing and displacement transformations, and derive a general integral equation (Equation (12)), which describes nonpertubatively the joint interaction of an electron with the atomic nucleus and with the quantized radiation field. In Section 3 we shall apply the derived integral equation, and discuss the general structure of the transition matrix elements, which may describe various multiphoton processes, like high-harmonic production, radiative electron scattering. A special emphasis will be put on the connection of the quantum and semiclassical descriptions. We shall show a new exact treatment of this question, based on the von Neumann lattice coherent states. Interestingly, our approach describes the traces of depletion, even in the semiclassical limit. In Section 4 we derive the quantized strong-field Kramers–Heisenberg formula for high-intensity light scattering, which seems to be a proper theoretical tool for analyzing the possible role of photon statistics in the HHG process. Section 5 closes the paper by summarizing the main results and conclusions. We have attempted to have the present paper possibly self-contained. That is why two appendices have been included, which, for definiteness, also summarize the basic notions, notations and the necessary details of the calculations. Besides, these appendices contain some additional background information.

## 2. General Formulation—Nonperturbative Treatment of the Interaction of a Bound or Free Electron with the whole Quantized Radiation Field

In the present section we are introducing a first-principle theoretical model based on nonrelativistic quantum electrodynamics, with the help of which we shall analyse the high-order quantum transitions of the system “single active electron + (strong) laser mode(s) + mode(s) of the scattered radiation”. The Schrödinger equation of the system under discussion reads

$$\left[\frac{1}{2m}\left(\hat{\boldsymbol{p}}+\frac{e}{c}\hat{\boldsymbol{A}}\right)^2+V(\boldsymbol{r})+\hat{H}_{rad}\right]|\Psi(t)\rangle=i\hbar\partial_t|\Psi(t)\rangle\,, \tag{1}$$

where $V(\boldsymbol{r})$ is an arbitrary binding (or scattering) potential of the electron. The vector potential and energy operator of the complete radiation field is given as sums of contributions of the individual modes,

$$\hat{\boldsymbol{A}}=\sum_{\boldsymbol{k},s}\boldsymbol{A}_{\boldsymbol{k},s}(\boldsymbol{r})=\sum_{\boldsymbol{k},s}c\sqrt{2\pi\hbar/\omega_k L^3}\,\boldsymbol{\varepsilon}_{\boldsymbol{k},s}(a_{\boldsymbol{k},s}\mathrm{e}^{i\boldsymbol{k}\cdot\boldsymbol{r}}+a^{+}_{\boldsymbol{k},s}\mathrm{e}^{-i\boldsymbol{k}\cdot\boldsymbol{r}})\,, \tag{2}$$

$$\hat{H}_{rad}=\sum_{\boldsymbol{k},s}\hbar\omega_k(a^{+}_{\boldsymbol{k},s}a_{\boldsymbol{k},s}+\tfrac{1}{2})\,. \tag{3}$$

The elementary charge and the mass of the electron are denoted by $e$ and $m$, respectively, $c$ is the velocity of light in vacuum, and $\hbar$ is Planck’s constant divided by $2\pi$. The radiation is described by linearly polarized plane waves, which satisfy the vacuum dispersion relation $\omega_k=c|\boldsymbol{k}|$. The polarization vectors $\boldsymbol{\varepsilon}_{\boldsymbol{k},1}$, $\boldsymbol{\varepsilon}_{\boldsymbol{k},2}$ and the propagation vector $\boldsymbol{k}$ form a right system of orthogonal vectors for each mode. The quantized amplitudes (photon absorption and emission operators) satisfy the commutation relations $[a_{\boldsymbol{k},s},a^{+}_{\boldsymbol{k}',s'}]=\delta_{\boldsymbol{k},\boldsymbol{k}'}\delta_{s,s'}$ (all other type of commutators are zero). The formal details of the quantization of the radiation field can be found in any textbook of quantum theory [92,93] or quantum optics [94–96], as well as in review papers, like [97]. Anyway, for definiteness, in Appendix A we summarize some basic notions, notations and mathematical relations, which will be used in the main text. In dipole approximation the $\boldsymbol{r}-$dependence of the vector potential is neglected, more precisely, $\boldsymbol{r}$ can be replaced by the position of the atomic nucleus $\boldsymbol{r}_n$. The resulting constant phase factors $\exp(i\boldsymbol{k}\cdot\boldsymbol{r}_n)$ are mere parameters, for simplicity, they will not be shown. Since these phase factors do not affect the commutation relations, they can be thougth of being incorporated to the amplitude operators. At any stage of the discussion, we can re-display them, simply by the replacement $a_{\boldsymbol{k},s}\to a_{\boldsymbol{k},s}\exp(i\boldsymbol{k}\cdot\boldsymbol{r}_n)$. By expanding the square on the left hand side of Equation (1) we arrive at the explicit expressions for the electron-radiation interactions terms,

$$\left[\frac{\hat{\boldsymbol{p}}^2}{2m}+V(\boldsymbol{r})+\hat{K}+\hat{M}\right]|\Psi(t)\rangle=i\hbar\partial_t|\Psi(t)\rangle\,, \tag{4}$$

where

$$\hat{K}=\sum\nolimits_k\hat{K}_k\,,\quad \hat{K}_k=(e/mc)\hat{\boldsymbol{p}}\cdot\hat{\boldsymbol{A}}_k+(e^2/2mc^2)\hat{A}_k^2+\hbar\omega_k(a_k^+a_k+\tfrac{1}{2})\,, \tag{5}$$

$$\hat{M}=\sum\nolimits_k\sum\nolimits_{k'<k}\hat{M}_{k,k'}\,,\quad \hat{M}_{k,k'}=(e^2/mc^2)\hat{\boldsymbol{A}}_k\cdot\hat{\boldsymbol{A}}_{k'}\,. \tag{6}$$

In Equations (5,6) we have introduced a condensed notation $k\equiv(\boldsymbol{k},s)$ for the mode index. The $\hat{K}_k$ contain the "$\boldsymbol{p}\cdot\boldsymbol{A}$", "$\boldsymbol{A}^2$" interaction terms, the energy of a mode, and $\hat{M}_{k,k'}$ describe the *direct* mode-mode coupling. It is possible to decompose the original Hamiltonian in Equation (1) in a different way, where we distinguish the (initially) occupied modes, which represent the incoming radiation (e.g., the strong laser field). This leads to

$$\left[\frac{\hat{\boldsymbol{p}}^2}{2m}+V(\boldsymbol{r})+\hat{K}^{(L)}+\hat{M}^{(L)}+\hat{N}^{(H)}+\hat{H}_{rad}^{(H)}\right]|\Psi(t)\rangle=i\hbar\partial_t|\Psi(t)\rangle\,, \tag{7}$$

where

$$\hat{K}^{(L)}=\sum\nolimits_k\hat{K}_k^{(L)}\,,\quad \hat{K}_k^{(L)}=[(e/mc)\hat{\boldsymbol{p}}\cdot\hat{\boldsymbol{A}}_k+(e^2/2mc^2)\hat{A}_k^2+\hbar\omega_k(a_k^+a_k+\tfrac{1}{2})]^{(L)}\,, \tag{8}$$

$$\hat{M}^{(L)}=\frac{e^2}{mc^2}\sum\nolimits_k\sum\nolimits_{k'<k}\hat{\boldsymbol{A}}_k^{(L)}\cdot\hat{\boldsymbol{A}}_{k'}^{(L)}\,, \tag{9}$$

$$\hat{N}^{(H)}=\frac{e}{mc}\left(\hat{\boldsymbol{p}}+\frac{e}{c}\hat{\boldsymbol{A}}^{(L)}\right)\cdot\hat{\boldsymbol{A}}^{(H)}+\frac{e^2}{2mc^2}\left(\hat{\boldsymbol{A}}^{(H)}\right)^2. \tag{10}$$

In Equations (8-10) the superscript (L) refers to the laser modes (L-modes), and (H) refers to the weak modes (H-modes, e.g., the potential harmonics). The interaction terms of $N^{(H)}$ can describe spontaneous emission, or first and second order absorption and stimulated emission of the H-modes. In both decompositions (5,6) and (8-10) the vector potential and the radiation Hamiltonian are the same, of course, as that, given in (2) and (3), respectively. It is clear that at this stage the distinction between L- and H-modes is quite arbitrary; mathematically, one may split the vector potential into two parts at will, and deal with one of them somehow nonperturbatively. For instance, the cross-term "$A^{(L)}.A^{(H)}$" is well known in the linear theory of light scattering, which dominates in the Thomson regime. In the forthcomings we shall diagonalize the "K" interaction terms in the dipole approximation. We note that the direct mode-mode coupling, $\hat{M}_{k,k'}$ $(k \neq k')$, could also be treated by introducing an orthogonal transformation of the amplitude operators, at the very beginning, however, we shall not discuss this more general method in the present paper. Our aim is to solve the basic Equation (4) (and (7)) in a nonperturbative way, in the sense that both the interaction with the quantized radiation field and with the atomic potential of the electron are jointly taken into account.

In the dipole approximation $\hat{K}_k$ (of either Equation (5) or Equation (8)) can be brought to the form

$$\hat{K}_k = \hbar\omega_k [\sqrt{\beta_k / m\hbar\omega_k}\,(\hat{\boldsymbol{p}}\cdot\boldsymbol{\varepsilon}_k)(a_k + a_k^+) + \tfrac{1}{2}\beta_k (a_k^2 + a_k^{+2}) + (1+\beta_k)(a_k^+ a_k + \tfrac{1}{2})], \tag{11}$$

where we have defined the dimensionless parameter $\beta_k$,

$$\beta_k = 2\pi e^2 / mL^3\omega_k^2 = \omega_p^2 / 2\omega_k^2\,, \quad \omega_p^2 = 4\pi n_e e^2 / m\,, \quad n_e = 1/L^3\,. \tag{12}$$

(Since the general case contains the special one, with $K^{(L)}$, we shall not show the optional superscript (L).) In Equation (12) we have formally introduced the plasma frequency $\omega_p$ associated to the electron density $1/L^3$. We note that in a collective Hamiltonian of $N$ background free electrons (stemming from photo-ionization, for instance) interacting with the quantized radiation field, the density $N/L^3$ automatically appears. We also note that, in the context of a recent experiment, a numerical estimate of $L^3$ and the plasma frequency can be found in Appendix B, after Equations (A38) and (A42), respectively.

Before we go on to the diagonalization procedure, let us make a few remarks on the last term on the right hand side of Equation (11), which has a nontrivial physical meaning. It contains the operator of the ponderomotive energy shift, well known in the semi-classical description. We note, however, that in the semi-classical description the ponderomotive potential appears as a c-number energy shift (AC Stark shift) of the electron's energy. Here, in the quantum optical description, it does not have an actual value; we may also say that this shift "belongs" to the radiation field, since it is operating on the Hilbert space of the quantized modes (though it contains the electron's parameters; charge and mass). As is shown in Appendix B (see Equation (A39)), the expectation value of this operator can be brought to the form

$$\sum_{\boldsymbol{k}} \hbar\omega_k \beta_k \left\langle (a_k^+ a_k + \tfrac{1}{2}) \right\rangle = \frac{1}{4}\mu^2 mc^2\,, \quad \mu = \frac{eF_\nu}{mc\omega} = 8.5\times 10^{-10}\sqrt{I}\lambda\,, \tag{13}$$

$$\frac{F_\nu^2}{8\pi} = \frac{\nu^2}{c^3}\Delta\Omega\Delta\nu(\bar{n}_\nu + \tfrac{1}{2})h\nu\,, \quad c\frac{F_\nu^2}{8\pi} = I_\nu \Delta\nu\,, \tag{14}$$

where $\nu = \omega / 2\pi$, and the summation has been extended to a (narrow) spectral range $\Delta\nu$ and solid angle $\Delta\Omega$ around some propagation direction. Equation (13) is the same as the usual expression for a single electron, where, moreover, we have introduced the dimensionless intensity parameter $\mu$ (dimensionless vector potential). In the numerical formula $I = I_\nu \Delta\nu$ denotes the intensity in Watt/cm$^2$, and the central wavelength $\lambda$ is measured in microns ($10^{-4} cm$). Equation (14) shows the interrelation of the mode density, energy density, spectral intensity $I_\nu$, and the average photon occupation number $\bar{n}_\nu$, and, at the same time, this equation *defines* the average field strength squared. We note that quite recently, the ponderomotive shift has been measured in the extreme ultraviolet regime. We also note that in Appendix B we have given an estimate on the number of relevant modes of the incoming laser field in the interaction volume of the high-harmonic generation, by using the parameters of another recent experiment (see considerations after Equation (A38)). According to this estimate the number of modes of an *incoming* laser field (spatial and temporal) can be very large (thousands of spatial modes), depending on the focusing and pulse length (ten-twenty longitudinal modes).

In order to eliminate the minimal coupling terms in the K operators in Equations (5,11) of the quantized radiation, we are using exactly the same algebraic method for the bound electrons, as the one used in our early papers [6,8] for free electrons. We can apply this method for an arbitrary countable set of modes. As a first step towards the solution, we diagonalize all the interaction terms $\hat{K}_k$ in Equation (4), by using the procedure described in details in Appendix B. We apply the Bogoliubov transformation (A29-31,A41), and the displacement transformation (A7,A43). Accordingly, the state of the system is transformed to a new state $\Phi_{SD}$

$$|\Psi(t)\rangle = \hat{S}\hat{D}|\Phi_{SD}(t)\rangle , \quad \hat{S} = \prod_k \hat{S}_k(\theta_k) , \quad \hat{D} = \prod_k \hat{D}_k(\sigma_k) , \tag{15}$$

where

$$\hat{S}_k(\theta_k) = \exp[-\tfrac{1}{2}\theta_k(a_k^{+2} - a_k^2)] , \quad \theta_k = \tfrac{1}{4}\log(1 + 2\beta_k) , \tag{16}$$

$$\hat{D}_k(\sigma_k) = \exp[\sigma_k(a_k^+ - a_k)] , \quad \sigma_k = -\sqrt{\beta_k / m\hbar\omega_k}\, e^{-3\theta_k}(\hat{\boldsymbol{p}} \cdot \boldsymbol{\varepsilon}_k) . \tag{17}$$

Then, in the equation for the new state the " $\boldsymbol{p} \cdot \boldsymbol{A}$ " and the " $\boldsymbol{A}^2$ " interaction terms are absent (but still we have the mode-mode coupling term)

$$\left[ \frac{\hat{\boldsymbol{p}}^2}{2m} + V(\boldsymbol{r} + \hat{\boldsymbol{\alpha}}) + \widetilde{H}_{rad} + \hat{D}^+ \hat{S}^+ \hat{M} \hat{S} \hat{D} \right] |\Phi_{SD}(t)\rangle = i\hbar\partial_t |\Phi_{SD}(t)\rangle . \tag{18}$$

Notice that the squeezing operator $\hat{S}$, and the displacement operator $\hat{D}$ do not depend on time in the Schrödinger picture we are using. As an expense of the performed elimination, a displacement of the electron's position in the atomic potential appears, which is the operator

$$\hat{\boldsymbol{\alpha}} = \sum_k \hat{\boldsymbol{\alpha}}_k \,, \quad \hat{\boldsymbol{\alpha}}_k = -i\sqrt{\tilde{\beta}_k \hbar / m\tilde{\omega}_k}\,\boldsymbol{\varepsilon}_k (a_k^+ - a_k) = \frac{e}{mc^2}\hat{\boldsymbol{Z}}_k \,, \tag{19}$$

being proportional with the Hertz potential $\hat{\boldsymbol{Z}}$ [98] of the quantized electromagnetic field (see Appendices A and B for details, in particular, Equations (A5,6) and (A44, A46)). In the semiclassical limit this displacement corresponds to the trajectory of a free electron under the effect of a time-dependent electric field. So the shifted atomic potential in (18) is the quantum analogon of the Kramers-Henneberger space-translated potential [62,63]. The "dressed energy operator" of the radiation field contains the "blue-shifted" frequencies $\tilde{\omega}_k$,

$$\tilde{H}_{rad} = \sum_k \hbar\tilde{\omega}_k (a_k^+ a_k + \tfrac{1}{2}) \,, \quad \tilde{\omega} = \omega\sqrt{1+2\beta} = \sqrt{c^2\,|\,\boldsymbol{k}\,|^2 + \omega_p^2} \,. \tag{20}$$

In a plasma environment the harmonics are born "blue-shifted", and these frequencies are inherited when they propagate out of the interaction region, which should be taken into account in accurate diagnostics (see additional remarks in Appendix B). In the single active electron approach the square root in the frequency expression can be approximated as $\tilde{\omega} \approx \omega(1+\beta)$, thus we get back to the ponderomotive term, which we have already discussed (see Equations (13,14) of the present section, and the corresponding part of Appendix B). So, we may write

$$\tilde{H}_{rad} = \hat{H}_{rad} + \hat{U}_{pond} \,, \quad \hat{U}_{pond} = \sum_k \hbar\omega_k \beta_k (a_k^+ a_k + \tfrac{1}{2}) \,. \tag{21}$$

Now we go over to the interaction picture, by transforming out the *total* unperturbed Hamiltonian (including the diagonal ponderomotive term),

$$|\Phi_{SD}(t)\rangle = \hat{U}_0(t)|\Phi(t)\rangle \,, \quad |\Psi(t)\rangle = \hat{S}\hat{D}\hat{U}_0(t)|\Phi(t)\rangle \,, \tag{22}$$

$$\hat{U}_0(t) = \exp[-(i/\hbar)(\hat{H}_{atom} + \hat{H}_{rad} + \hat{U}_{pond})t] \,, \quad \hat{H}_{atom} = \hat{\boldsymbol{p}}^2/2m + V(\boldsymbol{r}) \,. \tag{23}$$

Note that in the second equation of (22) the connection of the new state with the original one is also shown. From the point of view of our procedure, it is of outstanding importance, that before performing the transformation (first equation of Equation (22)), we have added and subtracted $V(\boldsymbol{r})$ in the effective Hamiltonian on the left hand side of Equation (18). In this way we were able to transform out the complete atomic Hamiltonian (not only the kinetic energy term, as is done in the strong field approximation, which then uses the Volkov states). This 'trick' has nothing to do with the quantized description of the radiation field. It also helps in the semiclassical theory, as we have shown for the first time long ago [36]. The resulting equation for the new state is

$$\hat{U}_0^+(t)\hat{D}^+\hat{S}^+[(V(\boldsymbol{r}) - \hat{S}\hat{D}V(\boldsymbol{r})\hat{D}^+\hat{S}^+) + \hat{M}]\hat{S}\hat{D}\hat{U}_0(t)|\Phi(t)\rangle = i\hbar\partial_t|\Phi(t)\rangle \,. \tag{24}$$

By integrating this equation with respect to time, we have

$$-\frac{i}{\hbar}\int_{t_0}^{t} dt'\hat{U}^+(t')[(V(\boldsymbol{r})-\hat{S}\hat{D}V(\boldsymbol{r})\hat{D}^+\hat{S}^+)+\hat{M}]\hat{U}(t')\left|\Phi(t')\right\rangle = \left|\Phi(t)\right\rangle - \left|\Phi(t_0)\right\rangle , \tag{25}$$

$$\hat{U}(t) = \hat{S}\hat{D}\hat{U}_0(t) , \tag{26}$$

where $\hat{S}$, $\hat{D}$ and $\hat{U}_0(t)$ have been defined in Equations (16,17) and (23), respectively. After having expressed the transformed state in terms of the original one (as is shown in the second equation of (22)), Equation (25) yields the integral equation,

$$\left|\Psi_0(t)\right\rangle = \hat{U}(t)\left\{\hat{U}^+(t_0)\left|\Psi_0\right\rangle - \frac{i}{\hbar}\int_{t_0}^{t} d\tau\hat{U}^+(\tau)[(V(\boldsymbol{r})-\hat{S}\hat{D}V(\boldsymbol{r})\hat{D}^+\hat{S}^+)+\hat{M}]\left|\Psi_0(\tau)\right\rangle\right\} . \tag{27}$$

This equation describes the evolution of some initial state $\left|\Psi_0\right\rangle$ of the complete system to the state $\left|\Psi_0(t)\right\rangle$. Equation (27) is one of the main results of the present paper, it describes nonpertubatively the joint interaction of an electron with the atomic nucleus and with the quantized radiation. On the basis of its first few iterations, the physical content of this new integral equation shall be discussed in some special cases in the next section. Here we just note that since the original Hamiltonian of the complete system is time-independent (see the left hand side of Equation (1) or (4)), the time-dependent transitions are usually described by assuming an adiabatic switching of the coupling of the subsystems. Thus, in the remote past (at some $t_0$) and in the remote future (at some $t_1$) the evolution operator defined in Equation (26) is simply $U_0$ (see (23), in which $U_{pond}$ should be supressed). This adiabatic coupling does not mean that we cannot consider short-pulse interactions, because the propagator $U(t)$ contains the interaction with the whole quantized radiation field. By superimposing the needed Fourier components, any sort of incoming pulses can be constructed, whose interaction with the electron is nonperturbatively incorporated in $U(t)$.

## 3. Features of the Strong-Field Quantum Optical Transitions. Connection to the Semi-Classical Approximation

In the present section we show some consequences of the integral Equation (27). In particular, we shall discuss the field-dressed atomic potential, which appears in the first iteration (see the integrand in Equation (32) below). At the same time, we shall take the opportunity to have a glance on the "microscopic background" of some well-known semi-classical results. We note that, if on the right hand side of (27) in the integrand we approximate the exact state with the initial ground state, and project to a final dressed free electron state, then we receive the quantized version of a Keldysh-type amplitude [9] for ionization, in the "**p.A** gauge" [12,13]. Of course, even in this approximation, here we would also obtain additional effects, like the back reaction of the electron on the field, and various radiation processes. We shall not discuss such processes further, they are out of the main subject of the present paper.

The first term on the right hand side of Equation (27) is the zeroth approximant,

$$\left|\Psi_0^{(0)}(t)\right\rangle = \hat{U}(t)\hat{U}^+(t_0)\left|\Psi_0\right\rangle . \tag{28}$$

This state may be considered as a quantum generalization (with respect to the radiation field) of the Coulomb-Volkov wave [12,60,61], if the initial state is a direct product of an atomic state and *some* state of the field. For example, let us take as initial state a product state of the electronic ground state $\left|\psi_g\right\rangle$, and such a field state, which is in vacuum state, except for one mode (say, the laser mode), being in a coherent state $\left|\alpha\right\rangle$. Then this state (adiabatically) becomes a multimode squeezed coherent state at time $t$. We use momentum representation for the ground state, and do not display unimportant phase factors (containing the zero-point energy). Then, we have the Fourier component of (28)

$$\left\langle \boldsymbol{p}\right|\Psi_0^{(0)}(t)\rangle = \left\langle \boldsymbol{p}\right|\psi_g\rangle e^{iI_p(t-t_0)/\hbar}\hat{S}\left|\alpha e^{-i\tilde{\omega}\cdot(t-t_0)} + \sigma(\boldsymbol{p})\right\rangle e^{i\sigma(\boldsymbol{p})\alpha\sin\tilde{\omega}(t-t_0)}\prod_k\left|\sigma_k(\boldsymbol{p})\right\rangle , \tag{29}$$

where $I_p$ is the ionization potential, the $\sigma_k(\boldsymbol{p})$ are now c-numbers, defined according to Equation (17), and the squeezing operator is as in (15,16). In obtaining (29), we have taken into account that a displacement operator generates coherent state from the vacuum (see (A8) in Appendix A). The $\sigma_k(\boldsymbol{p})$ represent a static bias of each field oscillators, so each mode of the radiation field is stretched like a spring. In order to obtain the wave function of the electron in the coordinate representation, we have to integrate with respect to the momentum parameter, which yield a highly entangled state. We have studied such kind of entanglement for Gaussian packets of quantized Volkov states in [77,78], as already mentioned in the Introduction. As the coupling is gradually switched-off, in the far future ($t \to t_1$), the bias vanishes (and the propagation becomes just a free evolution of the uncoupled system). As basis states we shall use the complete orthogonal set of the Coulomb-Volkov states, of the type (28),

$$\left|\Phi(E_r,\{n_k\};t)\right\rangle = U(t)\left|E_r\right\rangle\left|\{n_k\}\right\rangle , \quad \hat{U}(t) = \hat{S}\hat{D}\hat{U}_0(t) , \tag{30}$$

where $\left|E_r\right\rangle$ are the stationary states of the electron (which can be bound or continuum states), and $\left|\{n_k\}\right\rangle$ denotes the direct product of the photon number eigenstates of the quantized radiation field. The transition matrix elements between these states are physically acceptable, because in the remote past and future they have unperturbed propagation, owing to $U(t_{0,1}) = U_0(t_{0,1})$. The orthogonality relation of the states (30) (at an arbitrary time instant) is expressed as

$$\left\langle \Phi(E_{r'},\{n_k'\};t)\right|\Phi(E_r,\{n_k\};t)\rangle = \delta_{r,r'}\delta_{\{n_k'\},\{n_k\}} , \tag{31}$$

where $\delta_{r,r'}$ is a Kronecker delta for bound states, and for continuum states it is proportional with a Dirac delta $\delta(E_{r'} - E_r)$. The index $r$ is a shorthand also for other quantum numbers, like that of the angular momentum. By projecting to a final quantized Coulomb-Volkov state of the type given in Equation (30) to the first iteration of Equation (27), we receive the first order transition amplitude

$$T_{f,i}^{(1)} = -\frac{i}{\hbar}\int_{t_0}^{t} d\tau \langle E_{r'}, \{n'_k\} | \hat{U}_0^+(\tau)[(V(\boldsymbol{r}+\hat{\boldsymbol{\alpha}})-V(\boldsymbol{r})) + \hat{D}^+\hat{S}^+\hat{M}\hat{S}\hat{D}]\hat{U}_0(\tau) | E_r, \{n_k\}\rangle . \qquad (32)$$

Equation (32) shows that it is possible to take into account nonperturbatively the joint interaction with the quantized radiation field and the atomic potential, already in the first iteration. The initial state can be chosen, for instance, to the product of an excited state of the electron and a (highly occupied) photon number eigenstates of a mode of the incoming (strong) laser field, and all the other modes, are being in the vacuum state. At the end state this latter modes will be occupied with photons, among them are the high-harmonic components, too. The incoming (laser) field can also be taken as an arbitrary superposition (like a coherent state or a squeezed state), and the complete transition matrix element will be a sum, weighted by the occupation amplitudes of the superposition. Asymptotically, the time-integral in (32) delivers a delta function, which secures the energy conservation,

$$E_f + \sum_{k'} m_{k'}\hbar\tilde{\omega}_{k'} = E_i + \sum_k n_k \hbar\tilde{\omega}_k , \qquad (33)$$

where $E_i$, $E_f$ are the initial and final energies of the electron, respectively, $m_{k'}$, $n_k$ are the corresponding photon numbers, and $\tilde{\omega}_k$ are the dressed frequencies (see Equation (20)). The non-perturbative transition amplitudes in Equation (32) can describe various high- (or low-) order processes, which are allowed by the energy conservation. An example is spontaneous emission during laser-assisted scattering. In this context see Ref. [89] (and references therein), where the laser was treated as an external field. We do not have space to deal with details of this type of amplitudes, except for the space-translated potential in it, whose appearance is straightforward in our formalism. This potential plays a central role in our forthcoming discussions. Besides, the study of this potential gives us an opportunity to derive new, and quite general relations, which connect the quantum optical and the semi-classical descriptions in strong-field physics.

The amplitudes (32) are determined by the matrix elements of the quantized space-translated potential $V_{\hat{\alpha}}$, which we shall call “multiphoton effective potential”,

$$V_{\hat{\alpha}} = V(\boldsymbol{r}+\hat{\boldsymbol{\alpha}}) - V(\boldsymbol{r}), \quad V(\boldsymbol{r}+\hat{\boldsymbol{\alpha}}) = \int d^3q \tilde{V}(\boldsymbol{q}) \exp[(i/\hbar)\boldsymbol{q}\cdot(\boldsymbol{r}+\hat{\boldsymbol{\alpha}})] . \qquad (34)$$

The Fourier representation of the space-translated part is also displayed in Equation (34). An illustration of this potential in real space is shown in Figure 1, more precisely, its semiclassical version with a c-number oscillating displacement is plotted. By taking Equation (19) for $\hat{\boldsymbol{\alpha}}$ into account, and remembering the general definition of the displacement operator (see Appendix A), we can rewrite the exponent in (34) as a product of displacement operators. We write out in (34) the expression of the corresponding phase factor for one mode,

$$\exp[(i/\hbar)\boldsymbol{q}\cdot\hat{\boldsymbol{\alpha}}_k] = \hat{D}[\gamma_k(\boldsymbol{q})], \quad \hat{D}[\gamma_k(\boldsymbol{q})] = \exp[\gamma_k(\boldsymbol{q})(a_k^+ - a_k)] , \qquad (35)$$

$$\gamma_k(\boldsymbol{q}) = \sqrt{\tilde{\beta}_k / \hbar m \tilde{\omega}_k}\,(\boldsymbol{q}\cdot\boldsymbol{\varepsilon}_k) . \qquad (36)$$

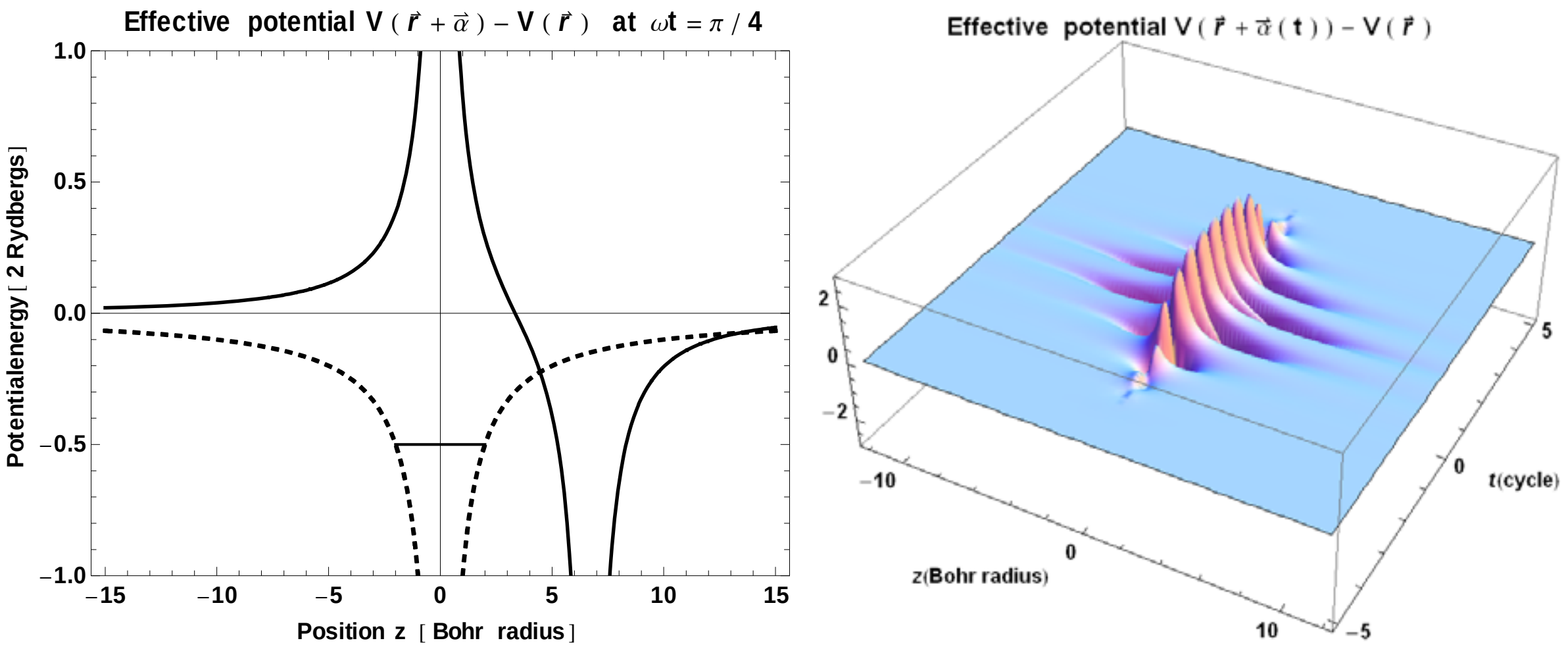

**(a)** **(b)**

**Figure 1.** This figure shows the multiphoton effective potential $V_{\hat{\alpha}} = V(\boldsymbol{r}+\boldsymbol{\alpha}) - V(\boldsymbol{r})$, given by Equation (34), for $V(\boldsymbol{r}) = -e^2/r$, in the semi-classical limit, where $\hat{\boldsymbol{\alpha}}$ is replaced by a c-number $\boldsymbol{\alpha}(t)$ (see equation (A47) in Appendix B): (**a**) the continuous line is the effective potential along the polarization direction, and the dashed line represent the original Coulomb potential. By using (A47), we have taken $\lambda = 0.8\times10^{-4}\,cm$, and $I = 3\times10^{13}\,W/cm^2$, in which case the amplitude of the free electron oscillation (or Hertz potential) is 9 times larger than the Bohr radius ($a_0 = \hbar^2/me^2 = 5\times10^{-9}\,cm$). We have taken the value $\rho = 0.001$ for the cylindrical coordinate; (**b**) this 3D figure illustrates the temporal evolution of the effective potential in Heisenberg picture. We have taken one quarter of the former intensity and assumed a Gaussian envelope $\exp(-t^2/2)$, which corresponds to 2.3 cyles full temporal width at half maximum. For a better visualization, we have chosen $\rho = 0.32$ for the cylindrical coordinate. This figure illustrates, that the effective potential vanishes before and after the external pulse, in contrast to the usual Kramers-Henneberger space translated potential $V(\boldsymbol{r}+\boldsymbol{\alpha}(t))$ [62,63].

Accordingly, the matrix elements of the multiphoton transitions are governed by the matrix elements of the above displacement operators (see Equations(A15,A16) in Appendix A),

$$d_{k,n} \equiv \langle k|\hat{D}(\gamma)|n\rangle = \begin{cases} (n!/k!)^{1/2}\gamma^{k-n}L_n^{k-n}(|\gamma|^2)\mathrm{e}^{-|\gamma|^2/2}, & (k \ge n) \\ (k!/n!)^{1/2}(-\gamma^*)^{n-k}L_k^{n-k}(|\gamma|^2)\mathrm{e}^{-|\gamma|^2/2}, & (0 \le k < n) \end{cases}. \tag{37}$$

In order to derive the semi-classical version of the quantum matrix elements $d_{k,n}$ defined in (37), we construct the superposition, which is weighted by coherent state occupation amplitudes, by keeping the order of the transition fixed (see Equations (A17,A18) in Appendix A). We have

$$\sum_{n=l}^{\infty}\frac{(\beta^*)^{n-l}}{\sqrt{(n-l)!}}\mathrm{e}^{-\frac{1}{2}|\beta|^2}\langle n-l|\hat{D}(\gamma)|n\rangle\frac{\alpha^n}{\sqrt{(n)!}}\mathrm{e}^{-\frac{1}{2}|\alpha|^2}, \tag{38}$$

where $\alpha$ and $\beta$ are some initial and final complex amplitude parameters, say, of a laser mode, and $l$ is a fixed number of absorbed photons. As is shown in Equation (A19), the summation can be performed analytically. We have derived a general formula for the coherent superpositions of the type Equation (38) for both $m-$photon emission and absorption processes, and found the unifying formula (see (A21)),

$$e^{-|\gamma|^2/2} e^{-\frac{1}{2}|\beta|^2-\frac{1}{2}|\alpha|^2+\beta^*\alpha}\left(e^{i\chi}\sqrt{\beta^*/\alpha}\right)^m J_m(2|\gamma|\sqrt{\beta^*\alpha}) \quad (m=0,\pm 1,\pm 2,\dots), \tag{39}$$

where $\chi = \arg(\gamma)$ (this phase can be non-zero, e.g., if we consider a circularly polarized mode). $J_m(z)$ is an ordinary Bessel function of first kind of order $m$. If $\beta = \alpha$, then Equation (39) exactly reduce to the a Jacobi-Anger expansion of the well-known Volkov state, in addition, we identify the energy density of a single laser mode with $\hbar\omega|\alpha|^2/L^3$ (see Equation (A22) in Appendix A). We would like to emphasize, that the formula (39) is exact; no approximations (like the the requirement of large occupation numbers, etc.) or additional assumptions have been used in deriving it. On the other hand, transition matrix elements between coherent states have to be considered with care. The initial state can in principle be any (normalized) state, e.g., a coherent state. However, according to the basic principles of quantum theory, transitions have a physical meaning for final states forming a complete orthogonal set (otherwise the probability interpretation of the alternatives is not possible). Fortunately, the discrete orthogonal system built up from the von Neumann lattice coherent states (see Appendix A) offers a legitimate basis set [80,82] for the final states. So, $\alpha$ and $\beta$ in Equation (39) can be considered as elements of the von Neumann lattice on the quantum phase space of the mode.

If $|\beta|<|\alpha|$, then one may think of the depletion of the particular mode. After all, as is seen from Equation (38), we are truly dealing with the depletion (decrease of the photon number in a mode) at the quantum level, so the choice of a smaller $|\beta|$ is intuitively justified. We note that recently there have been several detailed studies of the depletion effect in high-harmonic generation, including successful experimental works [84–86]. It is remarkable, that the general expression Equation (39) can also be expressed as a Fourier coefficient of an exponential with an oscillating trajectory (or Hertz potential), however, this "trajectory" contains an imaginary part, and also depends on the absolute phase (CEP),

$$t^k J_k(z) = \frac{1}{2\pi}\int_0^{2\pi} d\phi e^{-ik\phi}\exp(\gamma\beta^* e^{i\phi} - \gamma^*\alpha e^{-i\phi}), \quad t = e^{i\chi}\sqrt{\beta^*/\alpha}, \quad z = 2|\gamma|\sqrt{\beta^*\alpha}, \tag{40}$$

$$\exp\{i|\gamma|[(+i)(|\alpha|-|\beta|)\cos(\phi+\chi-\varphi)] + (|\beta|+|\alpha|)\sin(\phi+\chi-\varphi)]\}. \tag{41}$$

In Equation (41) we display the exponential in the integrand in (40), in the special case when the (final) phase of $\beta$ coincides with that of $\alpha = |\alpha|\exp(i\varphi)$. The first term in the bracket in the exponential is an imaginary contribution to a "trajectory" (since, the variable $\phi$ may be interpreted as $\omega \cdot t$, as showing up in the semiclassical theory). If $|\beta| = |\alpha|$, then this extra term is zero, and the second term exactly coincides with the oscillating phase of a Volkov state (from which, by the Jacobi-Anger formula, the usual Bessel function is obtained). Here, as a result of the *depletion*, "*imaginary satellite trajectories*" appear, as we may call them. This complex “trajectories” have nothing to do with the purely mathematical tool used to find the stationary points on the complex plane, in integrating highly oscillatory functions [54]. It is clear that, the “trajectories” (41) can always be associated to a transition matrix element of the type (37). According to this picture, to each von Neumann lattice points, surrounding the initial $\alpha$ there belong also an imaginary satellite trajectory, as a result of the depletion. So the spreading of the Wigner function of the fundamental (laser) mode [80,82], due to depletion, may be connected with these imaginary trajectories. The classical correspondence outlined here, can be built up for any countable number of modes, however, the neglection of the direct mode-mode coupling sets a limit for the validity of this interpretation.

## 4. Quantum Optical Strong-Field Kramers-Heisenberg Formula for a Nonperturbative Treatment of High-Harmonic Generation

In the present section we apply a variant of the integral Equation (27), which can be derived by using the same diagonalization technique, however, we distinguish the “laser modes”. The interaction with the other modes will be treated as a perturbation. In the first step, we apply the Bogoliubov transformation for the complete field. Concerning the laser modes, this is an exact procedure, because we use dipole approximation for this modes. For the other modes it is an approximation, because we keep the spatial dependence, just to illustrate the possibility of multipole HHG. We use the decomposition, summarized in Equations (7-10), and diagonalize the minimal coupling term in (8), by applying the corresponding squeezing and displacement operators, defined in Equations (16,17). Then we can derive the integral equation

$$|\Psi_0(t)\rangle = \hat{U}_L(t)\left\{\hat{U}_L^+(t_0)|\Psi_0\rangle - \frac{i}{\hbar}\int_{t_0}^{t} d\tau \hat{U}_L^+(\tau)[(V(\boldsymbol{r}) - \hat{S}\hat{D}_L V(\boldsymbol{r})\hat{D}_L^+\hat{S}^+) + \hat{M}^{(L)} + \hat{N}^{(H)}]|\Psi_0(\tau)\rangle\right\}, \tag{42}$$

$$\hat{U}_L(t) = \hat{S}\hat{D}_L\hat{U}_0(t)\,, \quad \hat{U}_0(t) = \exp[-(i/\hbar)(\hat{H}_{atom} + \widetilde{H}_{rad})t]\,. \tag{43}$$

In Equation (42) $\hat{H}_{atom}$ is the complete Hamiltonian of the atomic electron, and $\widetilde{H}_{rad}$ is the dressed energy of the field, defined in Equation (20). We note that in $\hat{N}^{(H)}$, Equation (10), the “A$^2$-term” is neglected. Moreover, in order to further simplify the analysis, we shall consider only a single laser mode, so $\hat{M}^{(L)}$, Equation (9), cancels, thus we receive from Equation (42) the integral equation,

$$|\Psi_0(t)\rangle = \hat{U}_L(t)\left\{\hat{U}_L^+(t_0)|\Psi_0\rangle - \frac{i}{\hbar}\int_{t_0}^{t} d\tau \hat{U}_L^+(\tau)[(V(\boldsymbol{r}) - \hat{S}\hat{D}_L V(\boldsymbol{r})\hat{D}_L^+\hat{S}^+) + \hat{N}^{(LH)}]|\Psi_0(\tau)\rangle\right\}, \tag{44}$$

$$\hat{N}^{(LH)}(\boldsymbol{r}) = \frac{e}{mc}\left(\hat{\boldsymbol{p}} + \frac{e}{c}\hat{\boldsymbol{A}}^{(L)}\right)\cdot\hat{\boldsymbol{A}}^{(H)}(\boldsymbol{r})\,, \quad \hat{\boldsymbol{A}}^{(L)} = \sqrt{2\pi\hbar/\widetilde{\omega}L^3}\,\boldsymbol{\varepsilon}(a + a^+)\,, \tag{45}$$

$$\hat{\boldsymbol{A}}^{(H)}(\boldsymbol{r}) = c\sum\nolimits_{\boldsymbol{k}',s'}\sqrt{2\pi\hbar/\widetilde{\omega}_{k'}L^3}\,\boldsymbol{\varepsilon}_{\boldsymbol{k}',s'}(a_{\boldsymbol{k},s}\mathrm{e}^{i\boldsymbol{k}'\cdot\boldsymbol{r}} + a^+_{\boldsymbol{k}',s'}\mathrm{e}^{-i\boldsymbol{k}'\cdot\boldsymbol{r}})\,. \tag{46}$$

As basis states for the evaluation of (44) we use the zeroth-order dressed states, similar to the states in Equation (30),

$$\left|\Phi_L(E_r,\{n_k\};t)\right\rangle = \hat{S}\hat{D}_L U_0(t)\left|E_r\right\rangle\left|\{n_k\}\right\rangle , \tag{47}$$

$$\hat{D}_L = \hat{D}(\sigma) = \exp[\sigma(a^+ - a)] , \quad \sigma = -\sqrt{\tilde{\beta}/m\hbar\tilde{\omega}}(\hat{\boldsymbol{p}}\cdot\boldsymbol{\varepsilon}) . \tag{48}$$

The first iteration of Equation (44) results in the matix element

$$T_{f,i}^{(L)(1)} = -\frac{i}{\hbar}\int_{t_0}^{t} d\tau\left\langle E_{r'},\{n_k'\}\right|\hat{U}_0^+(\tau)[(V(\boldsymbol{r}+\hat{\boldsymbol{\alpha}})-V(\boldsymbol{r}))+\hat{N}^{(LH)}(\boldsymbol{r}+\hat{\boldsymbol{\alpha}})]\hat{U}_0(\tau)\left|E_r,\{n_k\}\right\rangle , \tag{49}$$

$$\hat{\boldsymbol{\alpha}} = \hat{\boldsymbol{\alpha}}_L = -i\sqrt{\frac{\tilde{\beta}\hbar}{m\tilde{\omega}}}\boldsymbol{\varepsilon}(a^+ - a) = \frac{e}{mc^2}\hat{\boldsymbol{Z}}_L . \tag{50}$$

In obtaining the last term on the right hand side in the integrand in Equation (49), we have made the approximation $\hat{D}_L^+\hat{S}^+\hat{N}^{(LH)}\hat{S}\hat{D}_L \approx \hat{D}_L^+\hat{N}^{(LH)}\hat{D}_L$, which causes an error of order less than $\omega_p^2/2(\omega')^2$, which is negligible at this place. The shift of the laser vector potential gives a contribution of order $L^{-3}$, which is also vanishing in comparison with $L^{-3/2}$. In this first approximation the multiphoton effective potential does not give a contribution to high harmonics, however the second term, the space-translated $\hat{N}^{(LH)}$ interaction, yields multipole (magnetic dipole, etc.) radiation of any order, which has also been analysed in a very detailed paper [83]. On the basis of Equations (45,46) and (49,50), the matrix elements of $\exp(-i\boldsymbol{k}'\cdot\hat{\boldsymbol{\alpha}})$ has the same structure as that in Equations (35-37); we just have to replace $\boldsymbol{q}$ with $\boldsymbol{k}'$, the latter being the wave vector of an outgoing harmonic. Concerning the semi-classical analysis of this term, the results of the previous section also apply here. For instance, one can derive the semiclassical scattering cross-section in case of a bound electron, a nonlinearity factor governed by the Bessel function multiplies the Fourier transform of the square of the ground state wave function. In the optical range the relative size of this non-resonant radiation should be very small, in comparison with the electric dipole high-harmonics generated nowadays routinely. On the other hand, this non-dipole term results also in even harmonics, and have different angular distributions, and polarization properties, so the separation of them may be possible, as has been emphasized in [83]. If the spatial dependence $\exp(-i\boldsymbol{k}'\cdot\boldsymbol{r})$ is neglected, then remaining (dipole) interaction does not contribute to high-harmonic generation, nor does the multiphoton effective potential in Equation (49), either. However, in the second iteration a well-working formula, the multiphoton Kramers–Heisenberg formula comes out, as we have shown long ago, in a semiclassical analysis [36].

By performing the second iteration of the integral equation in Equation (44), in the dipole approximation, we arrive at the formula for the transition amplitude between the dressed states of the type (47)

$$T_{1,0}^{(II)} = 2\pi i\delta\{\hbar\widetilde{\omega}' - [(n_0 - n_1)\hbar\widetilde{\omega} + E_0 - E_1]\}\frac{e}{mc}\sqrt{\frac{2\pi\hbar}{\omega' L^3}}(A_{1,0} + B_{1,0})\,, \tag{51}$$

$$A_{1,0} = \sum_r \frac{\langle\psi_1|(\hat{\boldsymbol{p}}\cdot\boldsymbol{\varepsilon}')|\psi_r\rangle\langle\psi_r|\langle n_1|V_{\hat{\alpha}}|n_0\rangle|\psi_0\rangle}{(E_r - E_0 - (n_0 - n_1)\hbar\widetilde{\omega} + i0)}\,, \tag{52}$$

$$B_{1,0} = \sum_r \frac{\langle\psi_1|\langle n_1|V_{\hat{\alpha}}|n_0\rangle|\psi_r\rangle\langle\psi_r|(\hat{\boldsymbol{p}}\cdot\boldsymbol{\varepsilon}')|\psi_0\rangle}{(E_r - E_0 + \hbar\widetilde{\omega}' + i0)}\,. \tag{53}$$

In Equations (51-53) $E_{0,1}$ and $\psi_{0,1}$ are initial and final energies and wave functions of the electron, respectively. The summation, indexed by the label $r$ runs through the intermediate states, for the continuum states this means an integration. The $n_{0,1}$ and $|n_{0,1}\rangle$ are the initial and final photon numbers and states, respectively, so, if the final electron state coincides with the initial one, then $(n_0 - n_1)-$th order high-harmonics are produced. The matrix elements (with respect to the photon degrees of freedom) $\langle n_1|V_{\hat{\alpha}}|n_0\rangle$ are responsible for this high-order transitions, where $V_{\hat{\alpha}}$ is the multiphoton effective potential, introduced in Equation (34). From the symmetry properties of this potential, one can prove that only odd harmonics are produced. In Figure 2 we show a space-time diagram of the two parts of the transition amplitude. As can be seen in Equation (52), the $A_{1,0}$ part has resonances in the continuum for all high-enough-order harmonics, which explains the well-know plateau in the harmonic spectrum. In the resonant terms $E_r$ should be replaced by $E_r - i\hbar\Gamma_r/2$, where $\Gamma_r$ is the inverse life time of these unstable states [92]. The magnitude of the transition probabilities also depend on the dipole matrix elements between the continuum states and the ground state (which determine also the strength of linear photoelectric transitions induced by a weak high-frequency field, however, here one has to use outgoing continuum waves [92]). As is well-known, thanks to the path-breaking work of Cooper [99], in a certain range of the energy spectrum of the photo-electrons the contributions of the dipole matrix elements are completely suppressed, leading to a minimum. Such (continuum–bound) dipole matrix elements automatically appear in our strong-field Kramers–Heisenberg formula (51), describing HHG (see the $A_{1,0}$ term, Equation (52)). So the appearance of the Cooper minimum in the HHG spectrum [100–103] can be understood without any additional considerations.

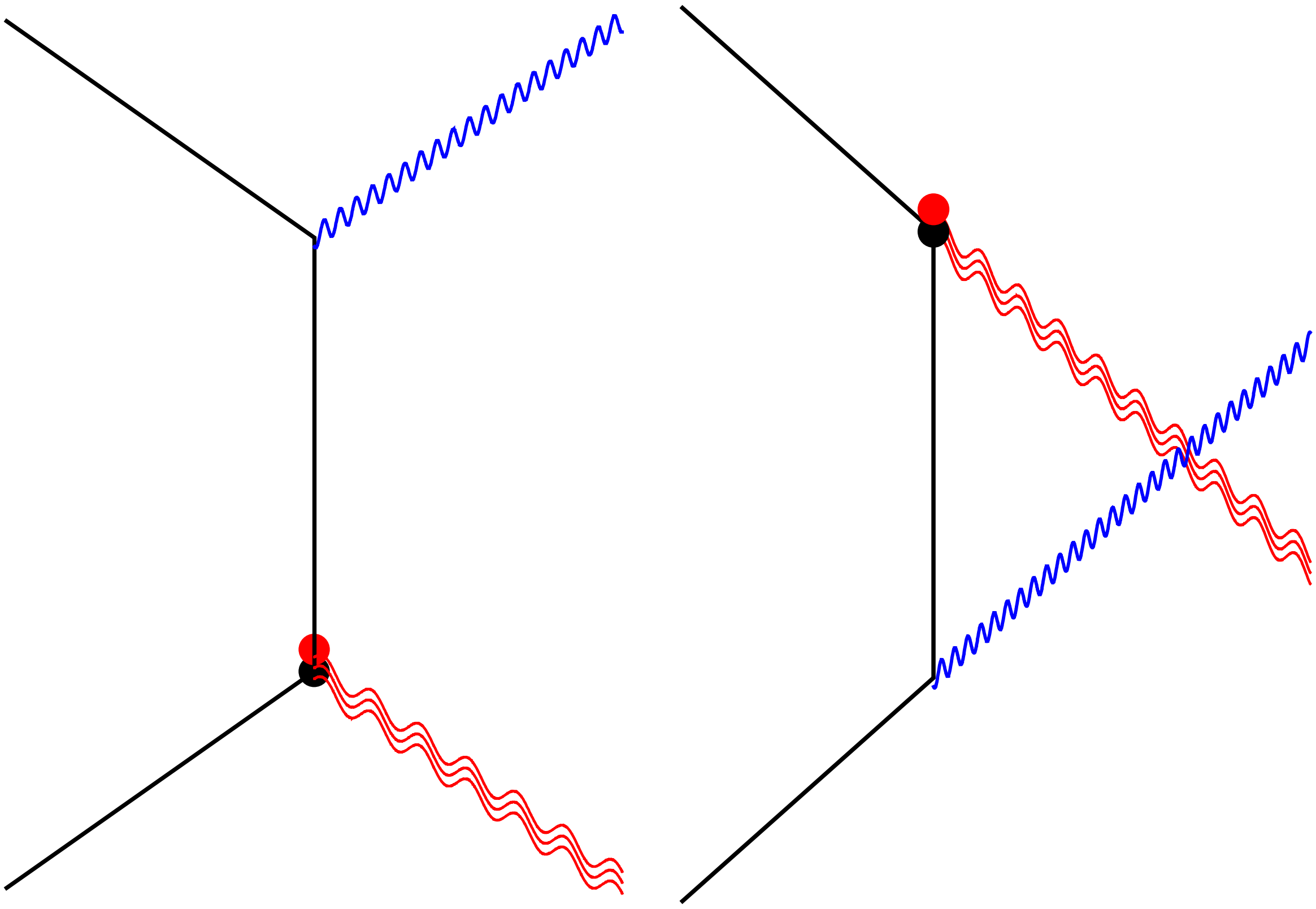

(**a**) (**b**)

**Figure 2.** Illustrates the two terms of the transition matrix elements (51) of the quantized strong-field Kramers–Heisenberg formula. The continuous black lines correspond to exact atomic electron initial, final and intermediate states. The red wavy lines symbolize the photon ansorption of the incoming strong laser field, and the blue wavy line refers to the outgoing harmonic radiation. The nonlinear interaction is condensed to the multiphoton effective potential $V_{\hat{\alpha}}$, which has been defined in Equation (34), and discussed in details in Section 3. In the figure this potential is symbolized by a pair of small discs, one being black (this refers to the original atomic potential), and the other disc is red-coloured (which refers to its dressing by the laser field); (**a**) The space-time diagram corresponding to the resonant term $A_{1,0}$, Equation (52); (**b**) is the space-time diagram corresponding to the matrix element $B_{1,0}$ of Equation (53).

## 5. Conclusions

In the present paper, in the frame of nonrelativistic quantumelectrodynamics, we have constructed a general theoretical model for nonperturbatively treating the interaction of an atomic electron with the quantized radiation field, which may contain high-intensity components. In the course of eliminating the minimal coupling terms of the quantized field, the interaction with the radiation is condensed to a space-translated “multiphoton effective potential”, which represent the vortices of multiphoton interaction. The connecting electron propagators represent exact (bound and continuum) atomic states of the electron, so this is a considerable improvement of the strong-field approximation. In Section 3 we have proved, that in our formalism the system "electron + (strong) laser mode(s) + mode(s) of the scattered radiation" has a natural basis set of entangled Coulomb-Volkov type states, whose photon part is in a multimode squeezed coherent state. It has been proved that a suitable coherent superposition of the transition matrix elements coincide with the semi-classical matrix elements, if we neglect the depletion of the field's coherent state. According to this result, our model may also serve as a proper "microscopic background" of various strong-field phenomena, like ATI or HHG. Besides, we have found that if the depletion of the laser mode is also taken into account, then complex trajectories can be associated to the generalized semi-classical transition amplitudes. The "imaginary satellite trajectory" is connected to the depletion. In Section 4 a quantum version of the strong-field Kramers-Heisenberg formula [36] has been derived, which may be a usable tool to discuss the effect of arbitrary photon statistics of the incoming field on the characteristics of the produced high harmonics. We have seen that the derived strong-field Kramers-Heisenberg formula inherently desribes the optional appearance of the Cooper minimum in the HHG spectrum. We note that the proposed formalism is also suitable for the description of radiation-assisted scattering in the presence of arbitrary quantized field, with the inclusion of the long-range effect in the continuum. The presented formalism may contribute to a more fundamental level of the theoretical description and understanding of photon-electron interactions in high-intensity radiation fields.

**Funding:** Support by the ELI-ALPS project is acknowledged. The ELI-ALPS project (GINOP-2.3.6-15-2015- 00001) is supported by the European Union and co-financed by the European Regional Development Fund.

**Acknowledgments:** The author dedicates the present paper to the memory of Fritz Ehlotzky (1929–2019), who had significantly contributed to the theory of strong-field phenomena over many decades.

## Appendix A. Basic Notions, Notations, and a Brief Summary on Coherent States and Squeezed States.

In free space, for the plane wave modes usually we prescribe periodic boundary conditions (i.e., $\boldsymbol{k} = 2\pi(m_x, m_y, m_z)/L$, where $m_x, m_y, m_z$ are integers) with a normalization volume $L^3$. The quantity $L^3\nu_k^2/c^3$ is the number of modes per unit solid angle per unit frequency range around the mode labelled by the index $k = (\boldsymbol{k}, s)$, where $\nu_k = \omega_k / 2\pi$ is the frequency. The quantized amplitudes (photon absorption and emission operators) of the radiation field satisfy the commutation relations $[a_{\boldsymbol{k},s}, a^+_{\boldsymbol{k}',s'}] = \delta_{\boldsymbol{k},\boldsymbol{k}'}\delta_{s,s'}$. All other commutators of the quantized amplitudes are zero. The "$n-$photon states" are eigenstates of the photon number operator for each mode $a^+_{\boldsymbol{k},s}a_{\boldsymbol{k},s}|n_{\boldsymbol{k},s}\rangle = n_{\boldsymbol{k},s}|n_{\boldsymbol{k},s}\rangle$ with $n_{\boldsymbol{k},s} = 0,1,2,...$ (We note that, in order that the equations have a more transparent appearance, we ocasionally suppress the mode index, if there is no risk of confusion.) As is well known, the photon absorption and emission are expressed with the equations $a|n\rangle = \sqrt{n}|n-1\rangle$ and $a^+|n\rangle = \sqrt{n+1}|n+1\rangle$, respectively, and any absorption operator annihilates the vacuum state (ground state) of the radiation field, i.e., $a|0\rangle = \vec{0}$, where $\vec{0}$ is the zero vector of the Hilbert space. The state vector $|\Psi(t)\rangle$ of the system in Equation (1) belongs to the product space $\mathscr{H}_e \otimes \mathscr{H}_{rad}$, where $\mathscr{H}_e$ is the electron's Hilbert space, and $\mathscr{H}_{rad} = \otimes_{\boldsymbol{k},s} \mathscr{H}_{\boldsymbol{k},s}$ is the Hilbert space of the whole radiation field. The operator part of a component $\hat{\boldsymbol{A}}_{\boldsymbol{k},s}(\boldsymbol{r})$ of the vector potential (as is writen in Schrödinger picture in Equation (2) of Section 2) can be brought to the alternative form, which has been used by Bloch and Nordsieck [70] in their path-breaking studies,

$$\hat{\boldsymbol{A}}_{\boldsymbol{k},s}(\boldsymbol{r}) = c\sqrt{2\pi\hbar / L^3\omega_k}\,\boldsymbol{\varepsilon}_{\boldsymbol{k},s}\sqrt{2}(P_{\boldsymbol{k},s}\cos\boldsymbol{k}\cdot\boldsymbol{r} + Q_{\boldsymbol{k},s}\sin\boldsymbol{k}\cdot\boldsymbol{r}), \tag{A1}$$

$$P_{\boldsymbol{k},s} = (a_{\boldsymbol{k},s} + a^+_{\boldsymbol{k},s})/\sqrt{2}, \quad Q_{\boldsymbol{k},s} = (a^+_{\boldsymbol{k},s} - a_{\boldsymbol{k},s})/i\sqrt{2}, \quad [P_{\boldsymbol{k},s}, Q_{\boldsymbol{k}',s'}] = -i\delta_{\boldsymbol{k},s;\boldsymbol{k}',s'}. \tag{A2}$$

We see that $P$ and $Q$ are hermitian and canonically conjugate operators, which satisfy the Heisenberg commutation relation. In quantum optics, essentially these are the so-called quadrature operators, usually labelled by $X$ and $Y$ [97], more precisely, $X = P/\sqrt{2}$ and $Y = -Q/\sqrt{2}$. In Heisenberg picture the vector potential and the electric field strength $\hat{\boldsymbol{E}}_{\boldsymbol{k},s}(\boldsymbol{r},t) = -\partial / c\partial t\hat{\boldsymbol{A}}_{\boldsymbol{k},s}(\boldsymbol{r},t)$ has an explicit time dependence,

$$\hat{\boldsymbol{A}}_{\boldsymbol{k},s}(\boldsymbol{r},t) = c\sqrt{2\pi\hbar / L^3\omega_k}\,\boldsymbol{\varepsilon}_{\boldsymbol{k},s}\sqrt{2}[P_{\boldsymbol{k},s}\cos(\omega_k t - \boldsymbol{k}\cdot\boldsymbol{r}) - Q_{\boldsymbol{k},s}\sin(\omega_k t - \boldsymbol{k}\cdot\boldsymbol{r})], \tag{A3}$$

$$\hat{\boldsymbol{E}}_{\boldsymbol{k},s}(\boldsymbol{r},t) = \sqrt{2\pi\hbar\omega_k / L^3}\,\boldsymbol{\varepsilon}_{\boldsymbol{k},s}\sqrt{2}[Q_{\boldsymbol{k},s}\cos(\omega_k t - \boldsymbol{k}\cdot\boldsymbol{r}) + P_{\boldsymbol{k},s}\sin(\omega_k t - \boldsymbol{k}\cdot\boldsymbol{r})]. \tag{A4}$$

In our formalism the time integral of the vector potential automatically appears, which is nothing else but the quantum analogon of the *electric Hertz potential* $\hat{\boldsymbol{Z}}_{\boldsymbol{k},s}(\boldsymbol{r},t)$ [98]. From $\hat{\boldsymbol{A}}_{\boldsymbol{k},s}(\boldsymbol{r},t) = +\partial / c\partial t\hat{\boldsymbol{Z}}_{\boldsymbol{k},s}(\boldsymbol{r},t)$ we have

$$\hat{\boldsymbol{Z}}_{\boldsymbol{k},s}(\boldsymbol{r},t) = (c^2/\omega_k)\sqrt{2\pi\hbar / L^3\omega_k}\,\boldsymbol{\varepsilon}_{\boldsymbol{k},s}\sqrt{2}[Q_{\boldsymbol{k},s}\cos(\omega_k t - \boldsymbol{k}\cdot\boldsymbol{r}) + P_{\boldsymbol{k},s}\sin(\omega_k t - \boldsymbol{k}\cdot\boldsymbol{r})]. \tag{A5}$$

In dipole approximation the factor $\sqrt{2}[...]$ in Equation (A5) simplifies to the expression of the type $-i(a^+\mathrm{e}^{i\omega t} - a\mathrm{e}^{-i\omega t})$ for each mode. For a classical electron moving in a classical radiation field $\boldsymbol{E}(t) = -\ddot{\boldsymbol{Z}}(t)/c^2$ (where the dot denotes derivative with respect to time) the solution of the Newton equation $-e\boldsymbol{E} = m\ddot{\boldsymbol{r}}$ is

$$\boldsymbol{r}(t) = \frac{e}{mc^2}\boldsymbol{Z}(t) \tag{A6}$$

(here we do not consider optional extra terms, which may come from initial values). Thus, the induced dipole $-e\boldsymbol{r} = r_0\boldsymbol{Z}$ is proportional to the electric Hertz vector, where $r_0 = e^2/mc^2$ is the classical electron radius.

Next, we summarize the main properties of the displacement operator $\hat{D}(\alpha)$ and the coherent states $|\alpha\rangle$, which we need in the main text. The unitary transformation generated by $\hat{D}(\alpha)$ displaces the amplitude operators,

$$\hat{D}(\alpha) = \exp(\alpha a^+ - \alpha^* a)\,,\quad \hat{D}^+(\alpha) a \hat{D}(\alpha) = a + \alpha\,,\quad \hat{D}^+(\alpha) a^+ \hat{D}(\alpha) = a^+ + \alpha^*\,, \tag{A7}$$

where $\alpha$ is an arbitrary complex number. The displacement operator $\hat{D}(\alpha)$ is the creation operator of the coherent state $|\alpha\rangle$, the latter is defined as an eigenstate of the absorption operator,

$$a|\alpha\rangle = \alpha|\alpha\rangle\,,\quad \hat{D}(\alpha)|0\rangle = |\alpha\rangle\,,\quad \hat{D}(\alpha)\hat{D}(\beta) = \hat{D}(\alpha+\beta)\exp[i\,\mathrm{Im}(\alpha\beta^*)]\,. \tag{A8}$$

We have also displayed the multiplication rule (the third equation in Equation (A8)), from which it follows, at the same time, that $\hat{D}(\beta)$ transforms a coherent state $|\alpha\rangle$ to a new coherent state $|\alpha+\beta\rangle$ (with an additional phase factor $\exp[i\,\mathrm{Im}(\alpha\beta^*)]$). The expansion of $|\alpha\rangle$ in terms of the number states show that the photon number distribution in a coherent state is a Poisson distribution,

$$|\alpha\rangle = \exp(-\tfrac{1}{2}|\alpha|^2)\sum_{n=0}^{\infty}\frac{\alpha^n}{\sqrt{n!}}|n\rangle\,,\quad P_n = |\langle n|\alpha\rangle|^2 = \frac{|\alpha|^{2n}}{n!}\exp(-|\alpha|^2)\,. \tag{A9}$$

The coherent states have been introduced in quantum optics in the path-breaking works of Glauber [104,105]. We note that the coherent states of a harmonic oscillator was invented by Schrödinger [106] already in 1926. In the context of field theory, it has been shown [71,72,92] that an external (classical) current distribution generates coherent states (of photons, or bosons, in general) from the vacuum state. Earlier we have worked out this problem for the special case of an external current consisting of oscillating free electrons in a strong laser field. In this case the resulting high-harmonic components are in multimode (product) coherent states [73,74]. According to the multiplication rule in (A8), if the field is initially in a coherent sate, then a classical current transforms it to a new coherent state. The coherent states with different parameters are not orthogonal, but they form an (overcomplete) set in each mode's Hilbert space,

$$\langle\beta|\alpha\rangle = \exp(-\tfrac{1}{2}|\beta|^2 - \tfrac{1}{2}|\alpha|^2 + \beta^*\alpha)\,,\quad |\langle\beta|\alpha\rangle|^2 = \exp(-|\beta-\alpha|^2)\,. \tag{A10}$$

$$\frac{1}{\pi}\int_{\mathbb{C}} d^2\alpha|\alpha\rangle\langle\alpha| = 1\,,\quad d^2\alpha = d(\mathrm{Re}\,\alpha)d(\mathrm{Im}\,\alpha)\,, \tag{A11}$$

where the integration goes over the whole complex plane (phase space). The expectation value of a component of the electric field strength (A4) in a coherent state $|\alpha_{\boldsymbol{k},s}\rangle$ exactly looks like a classical monochromatic plane wave,

$$\langle\hat{\boldsymbol{E}}_{\boldsymbol{k},s}(\boldsymbol{r},t)\rangle = \sqrt{2\pi\hbar\omega_k/L^3}\,\boldsymbol{\varepsilon}_{\boldsymbol{k},s}2[(\mathrm{Im}\,\alpha_{\boldsymbol{k},s})\cos(\omega_k t - \boldsymbol{k}\cdot\boldsymbol{r}) + (\mathrm{Re}\,\alpha_{\boldsymbol{k},s})\sin(\omega_k t - \boldsymbol{k}\cdot\boldsymbol{r})]\,. \tag{A12}$$

Concerning the theory of nonclassical states, see e.g. [107–108]. The carrier-envelope phase difference (CEP) is contained in the complex parameter $\alpha_{\mathbf{k},s}$. Many *discrete* complete subsets of coherent states can be constructed, among them the so-called *von Neumann lattice coherent states* $\left|\alpha^{(n,m)}\right\rangle$ are of particular importance [109–111], where $\alpha^{(n,m)} = (n+im)\sqrt{\pi}$, and $n$ and $m$ are integer numbers. This means to divide the oscillator's phase-space to elemetary squares, whose area equals to just Planck's elementary quantum of action, $\Delta q \Delta p = h$. The modulus of the scalar product of two such lattice coherent states is

$$\left|\left\langle\alpha^{(n',m')}\right|\alpha^{(n,m)}\right\rangle\right| = \exp(-\pi\,|\,n'-n\,|^2)\exp(-\pi\,|\,m'-m\,|^2)\,, \tag{A13}$$

which clearly shows that already the next-next neighbours are 'practically orthogonal'. This property has been very useful in our quantum optical study of high-harmonic generation [80,82]. The completeness [110,111] of the von Neumann lattice coherent states is *approximately* expressed by the following double sum,

$$\sum_{n=-\infty}^{\infty}\sum_{m=-\infty}^{\infty}\left|\alpha^{(n,m)}\right\rangle\left\langle\alpha^{(n,m)}\right| \approx= 1\,. \tag{A14}$$

More precisely, by starting at any point $(n_0, m_0)$, we can uniquely label all the lattice coherent states $\left|\alpha^{(n,m)}\right\rangle$ by one single index $k = 0,1,2,...$, i.e. $\left|\alpha^{(n,m)}\right\rangle \leftrightarrow \left|\alpha_k\right\rangle$, where $\alpha^{(n,m)} = \sqrt{\pi}(n+im)$. Then, the true completeness relation can be written as

$$\sum_{k=0}^{\infty}\sum_{l=0}^{\infty}\left|\alpha_k\right\rangle A_{kl}\left\langle\alpha_l\right| = 1\,, \tag{A14a}$$

where the matrix $A = \{A_{kl}\}$ is the inverse of the so-called Gram matrix, $M = \{M_{kl}\}$, which is defined as $M_{kl} = \left\langle\alpha_k\right|\alpha_m\rangle$. The "technical advantage" of using von Neumann lattice coherent states, rest on the property that any finite set of them is a linearly independent set. By orthogonalizing such sets, we receive a basis set, being 'legitimate' from the point of view of quantum measurement theory. This means that transition matrix elements between this orthogonal states may have a physical meaning.

The matrix elements of a displacement operator $\hat{D}(\sigma)$ between photon number states has first been published by Bloch and Nordsieck [70]. We present this formula for a general complex (non-hermitian) $\sigma$, for emission of $m$ photons;

$$d_{n+m,n} = \left\langle n+m\right|\hat{D}(\sigma)\left|n\right\rangle = \sqrt{\frac{(n)!}{(n+m)!}}\sigma^m L_n^m(|\,\sigma\,|^2)\mathrm{e}^{-|\sigma|^2/2},\ \ (m \geq 0)\,, \tag{A15}$$

and, for absorption of $l$ photons;

$$d_{n-l,n} = \langle n-l|\hat{D}(\sigma)|n\rangle = \sqrt{\frac{(n-l)!}{(n)!}}(-\sigma^*)^l L_{n-l}^l(|\sigma|^2)\mathrm{e}^{-|\sigma|^2/2}, \ (l \ge 0), \tag{A16}$$

where $L_n^m(x)$ is an associated Laguerre polynomial [112]. If we take the coherent superpositions of these matrix elements (by keeping $m$ or $l$ fixed), with weights of the coherent states as is shown in (A9), we have

$$\sum_{n=0}^{\infty}\frac{(\beta^*)^{n+m}}{\sqrt{(n+m)!}}\mathrm{e}^{-\frac{1}{2}|\beta|^2}\frac{\alpha^n}{\sqrt{(n)!}}\mathrm{e}^{-\frac{1}{2}|\alpha|^2}\langle n+m|\hat{D}(\sigma)|n\rangle = \tag{A17}$$

$$= \mathrm{e}^{-\frac{1}{2}|\beta|^2}(\beta^*)^m\mathrm{e}^{-\frac{1}{2}|\alpha|^2}\sigma^m\mathrm{e}^{-|\sigma|^2/2}\sum_{n=0}^{\infty}\frac{(\beta^*\alpha)^n}{(n+m)!}L_n^m(|\sigma|^2), \ (m \ge 0). \tag{A18}$$

By using the following generating formula of the Laguerre polynomials (formula 8.975.3 of [112]),

$$J_\alpha(2\sqrt{xz})\mathrm{e}^z(xz)^{-\alpha/2} = \sum_{n=0}^{\infty}\frac{z^n}{\Gamma(n+\alpha+1)}L_n^\alpha(x) \quad [-1<\alpha]$$

we have the closed expression for the sum in (A17, A18) for emission

$$\mathrm{e}^{-|\sigma|^2/2}\mathrm{e}^{-\frac{1}{2}|\beta|^2-\frac{1}{2}|\alpha|^2+\beta^*\alpha}\left(\frac{\beta^*}{\alpha}\right)^{m/2}\mathrm{e}^{im\chi}J_m(2|\sigma|\sqrt{\beta^*\alpha}), \ (m \ge 0), \tag{A19}$$

where $J_m(z)$ is an ordinary Bessel function of first kind of order $m$, and $\chi = \arg(\sigma)$. A similar expression is valid for the coherent superposition of $l-$photon absorption matrix elements,

$$\mathrm{e}^{-|\sigma|^2/2}\mathrm{e}^{-\frac{1}{2}|\beta|^2-\frac{1}{2}|\alpha|^2+\beta^*\alpha}\left(\frac{\alpha}{\beta^*}\right)^{l/2}(-\mathrm{e}^{-i\chi})^l J_l(2|\sigma|\sqrt{\beta^*\alpha}), \ (l \ge 0). \tag{A20}$$

The unifying formula for the coherent superpositions of $m-$photon emission or absorption reads

$$\mathrm{e}^{-|\sigma|^2/2}\mathrm{e}^{-\frac{1}{2}|\beta|^2-\frac{1}{2}|\alpha|^2+\beta^*\alpha}\left(\mathrm{e}^{i\chi}\sqrt{\beta^*/\alpha}\right)^m J_m(2|\sigma|\sqrt{\beta^*\alpha}), \ (m = 0, \pm 1, \pm 2, \ldots). \tag{A21}$$

In case of $\beta = \alpha$, we receive for the transition amplitude (A21) a simpler form,

$$\mathrm{e}^{-|\sigma|^2/2}\mathrm{e}^{im(\chi-\varphi)}J_m(2|\alpha\sigma|), \quad \chi = \arg(\sigma), \quad \varphi = \arg(\alpha). \tag{A22}$$

Equations (A17-22) show how the semiclassical martix element is built up from the special coherent superposition of transition amplitudes $\langle n+m|\hat{D}(\sigma)|n\rangle$, all of them being associated to a fixed $m$ number of photon emission (absorption). This can be seen, if one takes into account the Jacobi-Anger formula [112], which is well known in strong-field theory,

$$\exp(iz\sin\phi) = \sum_{m=-\infty}^{\infty} J_m(z)\mathrm{e}^{im\phi}, \quad J_k(z) = \frac{1}{2\pi}\int_0^{2\pi} d\phi \mathrm{e}^{-ik\phi}\exp(iz\sin\phi). \tag{A23}$$

The usual sinusoidal phase modulation appears (coming from the $\boldsymbol{p}\cdot\boldsymbol{A}$ term in a Volkov state), from which the strengths of the Fourier components, $J_m(z)$, of the 'multiphoton side-bands' are derived. It is seen from Equation (A44) of Appendix B, that in momentum representation the argument of the Bessel function in (A22) (where $\bar{n} = |\alpha|^2$ is the average photon number) is just the corresponding semiclassical expression $(\mu/\hbar k)(\boldsymbol{p}\cdot\boldsymbol{\varepsilon})$, with $\mu = eF/mc\omega$ being the dimensionless intensity parameter of the laser field. The connection of the quantum and semi-classical matrix elements, represented (in the special case $\beta = \alpha$) by Equation (A22) has long been known, but the usual derivation relies on the asymptotic limit of the Laguerre polynomial (for a given $m$, and for a *given*) $n \to \infty$, which goes to infinity. Our method is more general, because one can treat incoherent sums of probabilities (the modulus square of the transition amplitude). The case $\beta \neq \alpha$ is of particular interest, because it may incorporates the depletion of the laser mode in classical terms. According to the generating formula [112]

$$\exp\left[\frac{z}{2}\left(t - \frac{1}{t}\right)\right] = \sum_{k=-\infty}^{\infty} t^k J_k(z), \quad |z| \le |t|, \tag{A24}$$

the general expression can be summed up. With $t = e^{i\chi}\sqrt{\beta^*/\alpha}$, and $z = 2|\sigma|\sqrt{\beta^*\alpha}$;

$$\sum_{k=-\infty}^{\infty} t^k \mathrm{e}^{ik\phi} J_k(z) = \exp(\sigma\beta^*\mathrm{e}^{i\phi} - \sigma^*\alpha\mathrm{e}^{-i\phi}), \tag{A25}$$

$$t^k J_k(z) = \frac{1}{2\pi}\int_0^{2\pi} d\phi \mathrm{e}^{-ik\phi}\exp(\sigma\beta^*\mathrm{e}^{i\phi} - \sigma^*\alpha\mathrm{e}^{-i\phi}). \tag{A26}$$

So, the semi-classical matrix element is a Fourier coefficient of the exponetiated "classical trajectory" (or Hertz potential)

$$\exp(\sigma\beta^*\mathrm{e}^{i\phi} - \sigma^*\alpha\mathrm{e}^{-i\phi}) =$$

$$= \exp\{|\sigma|[|\beta|\cos(\phi'-\varphi) - |\alpha|\cos(\phi'-\psi)] + i|\sigma|[|\beta|\sin(\phi'-\varphi) + |\alpha|\sin(\phi'-\psi)]\}, \tag{A27}$$

where $\phi' = \phi + \chi$, $\chi = \arg(\sigma)$, $\varphi = \arg(\beta)$, $\psi = \arg(\alpha)$. If the phase of $\alpha$ and $\beta$ are the same, then (A27) simplifies,

$$\exp\{|\sigma|[(|\beta| - |\alpha|)\cos(\phi'-\varphi)] + i(|\beta| + |\alpha|)\sin(\phi'-\varphi)]\}. \tag{A28}$$

At the end of the present appendix we present some properties of the squeezing operator $\hat{S}(\theta)$. In the special case when $\theta$ is a real parameter, the effect of $\hat{S}(\theta)$ on the amplitude operators is the following Bogoliubov transformation,

$$\hat{S}(\theta) = \exp[-\tfrac{1}{2}\theta(a^{+2} - a^2)] , \quad \text{(A29)}$$

$$a \to b = \hat{S}^+(\theta)a\hat{S}(\theta) = a\cosh\theta - a^+ \sinh\theta , \quad \text{(A30)}$$

$$a^+ \to b^+ = \hat{S}^+(\theta)a^+\hat{S}(\theta) = a^+ \cosh\theta - a\sinh\theta . \quad \text{(A31)}$$

The transformation of $a^+$ can simply be obtained by taking the adjoint of Equation (A30); this is a consequence of the unitarity of $\hat{S}(\theta)$. The “squeezing property” of this transformation [97] can be immediately obtained from (A30-31),

$$\hat{S}^+(\theta)X\hat{S}(\theta) = X\mathrm{e}^{-\theta} , \quad X = (a + a^+)/2 ; \quad \hat{S}^+(\theta)Y\hat{S}(\theta) = Y\mathrm{e}^{+\theta} , \quad Y = (a - a^+)/2i . \quad \text{(A32)}$$

If $\theta$ is positive, then the quadrature $X = P/\sqrt{2}$ is squeezed, and the quadrature $Y = -Q/\sqrt{2}$ is stretched.

## Appendix B. Elimination of the Minimal Coupling Interaction Terms.

In Section 2 we have encountered in Equation (11) quadratic expressions of the absorption and emission operators of the type

$$\hat{K}_k = \hbar\omega[\sqrt{\beta / m\hbar\omega}(\hat{\boldsymbol{p}} \cdot \boldsymbol{\varepsilon})(a + a^+) + \tfrac{1}{2}\beta(a^2 + a^{+2}) + (1 + \beta)(a^+ a + \tfrac{1}{2})] , \quad \text{(A33)}$$

(we have suppressed the mode index on the right hand side). The dimensionless parameters $\beta_k$ have been defined in Equation (12), for one mode we simply write here

$$\beta = 2\pi e^2 / mL^3\omega^2 = \omega_p^2 / 2\omega^2 , \quad \omega_p^2 = 4\pi n_e e^2 / m , \quad n_e = 1/L^3 . \quad \text{(A34)}$$

In Equation (A34) we have formally introduced the plasma frequency $\omega_p$ associated to the electron density $1/L^3$. The terms of the type $\hbar\omega\beta(a^+ a + 1/2)$ are in fact the operators of the ponderomotive energy shift. This can be shown in the following way. We sum up these terms in a (narrow) frequency range (around a frequency $\nu = \omega / 2\pi$ ), in a (small) solid angle around an average propagation vector,

$$\sum_{\boldsymbol{k}} \hbar\omega_k \beta_k (a_k^+ a_k + \tfrac{1}{2}) = \frac{L^3}{(2\pi)^3} \frac{2\pi e^2}{mL^3} \hbar \int d^3k \omega_k \frac{1}{\omega_k^2} (a_k^+ a_k + \tfrac{1}{2}) . \quad \text{(A35)}$$

By taking the average of the above expression, we have

$$\sum_{\boldsymbol{k}} \hbar\omega_k \beta_k \left\langle (a_k^+ a_k + \tfrac{1}{2}) \right\rangle = \left(\frac{e}{mc\omega}\right)^2 \frac{1}{4} mc^2 8\pi \int d\nu d\Omega(\boldsymbol{k}) \frac{\nu^2}{c^3} (\overline{n}_\nu + \tfrac{1}{2}) h\nu \,. \qquad \text{(A36)}$$

We associate a field strength parameter $F_\nu$ through the energy density formula

$$\frac{F_\nu^2}{8\pi} = \int d\nu d\Omega(\boldsymbol{k})(\overline{n} + \tfrac{1}{2})h\nu = \frac{\nu^2}{c^3}\Delta\Omega\Delta\nu(\overline{n}_\nu + \tfrac{1}{2})h\nu\,, \quad I_0 = c\frac{F_\nu^2}{8\pi} = \frac{\nu^2}{c^2}\Delta\Omega\Delta\nu(\overline{n}_\nu + \tfrac{1}{2})h\nu\,, \qquad \text{(A37)}$$

$$\lambda_0^2 I_0 \tau / \Delta\Omega = (\overline{n}_\nu + \tfrac{1}{2})h\nu\,. \qquad \text{(A38)}$$

To have an order of magnitude estimate for the gometrical factors and the number of modes, we make a brief detour, by using the parameters of a recent important experiment [113]. Let us consider an incoming laser beam, with diameter 6 mm (cross-sectional area being 0.28 cm$^2$, which is focused with a lens of 15 cm focal length to a 1.2 mm long gas jet [113]. The solid angle substended by the lens viewed from the target is $\Delta\Omega \approx 1.24 \times 10^{-3}$. For the cross-sectional area of the target (the jet) we take $f \approx 0.5 \times 10^{-2}$ cm$^2$. In the experiment [113] they have used a linearly polarized Ti: Sapphire laser pulse of 800 nm wavelength, and of duration $\tau \approx 35$ fs (the longitudinal size of the pulse is then about $10^{-3}$ cm). This corresponds to 13.5 cyles, so one may associate about 14 temporal (longitudinal) degrees of freedom. The number of (transverse) spatial degrees of freedom in general is $(f/\lambda^2)\Delta\Omega$, which equals to 967 in the present example. So the photons converging to the target are shared by a quite large number of modes (of order of ten thousand in this experiment). At this point we note that in the direct frequency comb technique (which has been extended to the UV region, by using high-harmonic radiation) the contribution of even one mode out of 20,000 (!) in a resonant interaction has already been measured [114].

Now we go back to discuss the expectation value of the ponderomotive shift. By taking (A36-38) into account, we finally have

$$\sum_{\boldsymbol{k}} \hbar\omega_k \beta_k \left\langle (a_k^+ a_k + \tfrac{1}{2}) \right\rangle = \left(\frac{eF_\nu}{mc\omega}\right)^2 \frac{1}{4} mc^2\,, \quad \frac{eF_\nu}{mc\omega} = \mu_0 = 8.5 \times 10^{-10}\sqrt{I_0}\lambda_0\,, \qquad \text{(A39)}$$

where we have introduced the averaged dimensionless intensity parameter $\mu_0$, and in the numerical formula $I_0$ denotes the intensity in Watt/cm$^2$, and the central wavelength $\lambda_0$ is measured in microns ($10^{-4}cm$). We note that Equation (B3a) connects the average photon number $\overline{n}_\nu$ of a spectral component with the energy density through the spectral mode density. It is well known that in a coherent state $|\alpha\rangle$ the photon number expectation value is $\overline{n} = |\alpha|^2$.

The diagonalization of the quadratic expression in (A33), can be performed by a consecutive application of the squeezing operator $\hat{S}(\theta)$ and the displacement operator $\hat{D}(\alpha)$, which have been defined in Appendix A, in Equations (A29) and (A7), respectively. Since the algebraic structure of the interaction is the same for all modes, this diagonalization procedure can be immediately applied for an arbitrary number of modes. We follow our earlier procedure, which has long been published in the studies on multiphoton Bremsstrahlung of a Schrödinger-electron [8] and on high-order harmonic generation (nonlinear Compton effect) on a Dirac-electron in a quantized radiation field [6]. First we apply a Bogoliubov transformation, whose generator is the unitary squeezing operator (A29), with a by now unknown parameter. After using (A30,A31), and having collected terms with the same combinations of the amplitudes, we realize that the off-diagonal quadratic terms drop out, if we require the following condition on the unknown parameter $\theta$,

$$\beta\cosh 2\theta-(1+\beta)\sinh 2\theta=0\,,\quad \sqrt{1+2\beta}=\mathrm{e}^{2\theta}\,,\quad \theta=\tfrac{1}{4}\log(1+2\beta)\,. \tag{A40}$$

As a result, we have

$$\hat{S}^{+}\hat{K}_{k}\hat{S}=\hbar\omega[\sqrt{\beta/m\hbar\omega}(\hat{\boldsymbol{p}}\cdot\boldsymbol{\varepsilon})(a+a^{+})\mathrm{e}^{-\theta}+(a^{+}a+\tfrac{1}{2})\mathrm{e}^{+2\theta}]\,,\quad \mathrm{e}^{2\theta}=\sqrt{1+2\beta}\,. \tag{A41}$$

It is remarkable, that in (A41) a "dressed frequency", $\widetilde{\omega}$ , can be introduced,

$$\widetilde{\omega}=\omega\mathrm{e}^{2\theta}=\omega\sqrt{1+2\beta}=\sqrt{\omega^{2}+\omega_{p}^{2}}\,,\qquad \widetilde{\omega}=\sqrt{c^{2}\,|\,\boldsymbol{k}\,|^{2}+\omega_{p}^{2}}\,. \tag{A42}$$

In a free electron gas of electron density $N/L^3$ Equation (A42) shows the dispersion relation of plasmons, propagating in an underdense plasma. For example, at the density $n_e=10^{18}/cm^3$ , the plasmon energy is $\hbar\omega_p=0.035eV$ , which is only slightly larger than two percent of the photon energy 1.5 eV. Then $\omega_p^2/2\omega^2=2\times10^{-4}$ , i.e., $\Delta\omega=2\times10^{-4}\omega$ , and this means $n\times(3\times10^{-4}eV)$ "blue-shift" of the $n-$th harmonics [115–117]. The remaining linear, off-diagonal term in (A41) can be transformed out by using the a displacement properties, shown in Equation (A7),

$$\hat{D}(\sigma)=\exp[\sigma(a^{+}-a)]\,,\quad \hat{D}^{+}(\sigma)a\hat{D}(\sigma)=a+\sigma\,,\quad \hat{D}^{+}(\sigma)a^{+}\hat{D}(\sigma)=a^{+}+\sigma\,, \tag{A43}$$

$$\sigma=-\sqrt{\beta/m\hbar\omega}\mathrm{e}^{-3\theta}(\hat{\boldsymbol{p}}\cdot\boldsymbol{\varepsilon})\,,\quad\text{or}\ \sigma=i\sqrt{\beta\hbar/m\omega}\mathrm{e}^{-3\theta}(\boldsymbol{\varepsilon}\cdot\nabla)\,. \tag{A44}$$

Note that $\sigma$ is a Hermitian operator acting on the electron's Hilbert space. By collecting terms, we end up with the diagonal expression for an arbitrary mode,

$$\hat{D}_{k}^{+}(\sigma_{k})\hat{S}_{k}^{+}(\theta_{k})\hat{K}_{k}\hat{S}_{k}(\theta_{k})\hat{D}_{k}(\sigma_{k})=\hbar\widetilde{\omega}_{k}(a_{k}^{+}a_{k}+\tfrac{1}{2}-\sigma_{k}^{2})\,, \tag{A45}$$

where we have again displayed the mode index $k\equiv(\boldsymbol{k},s)$. We also note that $\sqrt{\beta/\omega}\mathrm{e}^{-3\theta}$ in (A44) can also be written in the form $\sqrt{\widetilde{\beta}/\widetilde{\omega}}$ , if we use in the plasmon-dressed frequency $\widetilde{\omega}$ , introduced in (A42). The "blue-shift" is not the main subject of our present discussion, because, it would need a more detailed analysis in itself, thus we shall not keep track of the (small) change of the frequency due to this effect. The terms $\propto\sigma_k^2\propto(\hat{\boldsymbol{p}}\cdot\boldsymbol{\varepsilon}_k)^2$ can be incorporated to the kinetic energy of the electron in the complete Hamiltonian, Equation (3) in the main text. It can be shown that the sum of this kind of terms with respect to all the modes yields the contribution $(\hat{\boldsymbol{p}}^2/2m)(-m_{el}/m)$, where $m_{el}$ is the electromagnetic mass, which can be incorporated to the total mass, so we do not deal with these terms in the main text. As is seen in Equation (B5a), our displacement operator not merely displaces the quatized amplitudes, but it displaces also the electron's coordinate (along the polarization direction), because $\sigma$ contains the gradient ( $\hat{\boldsymbol{p}} = -i\hbar\nabla$ ). This means that the effect of the transformation generated by $\hat{D}(\sigma)$ shifts the argument of the potential $V(\boldsymbol{r})$ (see Equation (18) of Section 2),

$$\hat{D}^{+}(\sigma_k)V(\boldsymbol{r})\hat{D}(\sigma_k) = V(\boldsymbol{r}+\hat{\boldsymbol{\alpha}}_k)\,, \quad \hat{\boldsymbol{\alpha}}_k = -i\sqrt{\widetilde{\beta}_k\hbar / m\widetilde{\omega}_k}\,\boldsymbol{\varepsilon}_k(a_k^{+} - a_k)\,. \tag{A46}$$

The shifted potential is the quantum analogon of the classical space-translated potential [62,63]. In the Heisenberg picture the expectation value of the shift $\hat{\boldsymbol{\alpha}}_k(t)$ in some coherent state $|\alpha'_k\rangle$ is just the oscillating part of the classical trajectory of the electron,

$$\langle\alpha'_k|\hat{\boldsymbol{\alpha}}_k(t)|\alpha'_k\rangle = \frac{e}{mc^2}\langle\alpha'_k|\hat{\boldsymbol{Z}}_k(t)|\alpha'_k\rangle = \boldsymbol{\varepsilon}\alpha'_{k0}\sin\omega_k t\,, \quad \alpha'_{k0} = \mu_{k0}\lambda_{k0}/2\pi\,, \tag{A47}$$

where we have also shown the connection with the Hertz potential of the mode in dipole approximation (see Equations (A5,A6,A12) in Appendix A). According to the considerations leading to Equation (A39), the c-number parameter $\alpha'_{k0}$ (which is the oscillation amplitude of the electron) can be expressed as seen in (A47), where $\mu_{k0}$ is the dimensionless intensity parameter (also called "dimensionless vector potential") for a mode, as is given in Equation (A39).

We note that the above diagonalization procedure (i.e., the elimination of the off-diagonal terms of the quantized amplitudes in (A33)) can be extended to an arbitrary countable set of modes, one just needs to use the products of the corresponding squeeze operators $\hat{S}_k$ and displacement operators $\hat{D}_k$, as appears in the main text. In this general case, because of the mode-mode coupling, the $\sigma_k$ satisfy an infinite set of inhomogeneous coupled linear equations. However this coupling is proportional with $L^{-3}$, in contrast to the inhomogenity terms of order $L^{-3/2}$, thus the former can be left out of consideration, at least in the present study. Finally we note that in the dipole approximation we are using, the Doppler shift and the recoil energy are neglected, which, on the other hand, play an important role in the physics of free electron lasers [118–120].

**References.**

[1] Gordon, W. Der Comptoneffekt nach der Schrödingerschen Theorie. *Z. Für Phys.* **1927**, *40*, 117–133.

[2] Volkov, D.M. Über eine Klasse von Lösungen der Diracschen Gleichung. *Z. Für Phys.* **1935**, *94*, 250–260.

[3] Brown, L.S.; Kibble, T.W.B. Interaction of intense laser beams with electrons. *Phys. Rev.* **1964**, *133*, A705–A719.

[4] Goldman, I.I. Intensity effects in Compton scattering. *Phys. Lett.* **1964**, *8*, 103–106.

[5] Ritus, V.I.; Nikishov, A.I. Quantum electrodynamics of phenomena in intense fields. *Works Lebedev Phys. Inst.* **1979**, *111*, 5–278. English translation in *J. Sov. Laser Res*. **1985**, *6*, 619–728. (In Russian).

[6] Bergou, J.; Varró, S. Nonlinear scattering processes in the presence of a quantized radiation field: II. Relativistic treatment. *J. Phys. A Math. Gen.* **1981**, *14*, 2281–2303.

[7] Bersons, I.Y. Electron in the quantized field of a monochromatic electromagnetic wave. *Zhurnal Éksp. Teor. Fiz.* **1969**, *56*, 1627–1633. (English Translate) *Sov. Phys. JETP* **1969**, *29*, 871.

[8] Bergou, J.; Varró, S. Nonlinear scattering processes in the presence of a quantized radiation field: I. Nonrelativistic treatment. *J. Phys. A Math. Gen.* **1981**, *14*, 1469–1482.

[9] Keldish, L.V. Ionization in the field of a strong electromagnetic wave. *Zhurnal Éksp. Teor. Fiz.* **1964**, *47*, 1945–1957. [*Sov. Phys. JETP* **1965**, *20*, 1307–1314].

[10] Perelomov, A.M.; Popov, V.S.; Terent'ev, M.V. Ionizationof atoms in an alternating electric field: II. *Zhurnal Éksp. Teor. Fiz.* **1967**, *51*, 309. [*Sov. Phys. JETP* **1967**, *24*, 207].

[11] Popov, V.S. Method of imaginary time for periodical fields. *Yad. Fiz.* **1974**, *19*, 1140–1156. (In Russian).

[12] Faisal, F.H.M. Multiple absorption of laser photons by atoms. *J. Phys. B At. Mol. Phys.* **1973**, *6*, L89–L92.

[13] Reiss, H.R. Effect of an intense electromagnetic field on a weakly bound system. *Phys. Rev. A* **1980**, *22*, 1786–1813.

[13] Agostini, P.; Fabre, F.; Mainfray, G.; Petite, G.; Rahman, N.K. Free-Free Transitions Following Six-Photon Ionization of Xenon Atoms. *Phys. Rev. Lett.* **1979**, *42*, 1127.

[15] Chin, S.L.; Farkas, G.; Yergeau, F. Observation of Kr and Xe ions created by intense nanosecond $CO_2$ laser pulses. *J. Phys. B At. Mol. Opt. Phys.* **1983**, *16*, L223–L226.

[16] DiMauro, L.F.; Agostini, P. Ionization dynamics in stronglaser fields. *Adv. At. Mol. Opt. Phys*. **1995**, *35*, 79.

[17] Becker, W.; Grasbon, F.; Kopold, R.; Miloševic, D.B.; Paulus, G.G.; Walther, H. Above-threshold ionization: From classical features to quantum effects. *Adv. At. Mol. Opt. Phys*. **2002**, *48*, 35.

[18] Popov, V.S. Tunnel and multiphoton ionization of atoms and ions in a strong laser field; Keldish theory. *Sov. Phys. Uspekhi* **2004**, *47*, 855–885.

[19] Bunkin, F.V.; Fedorov, M.V. Bremsstrahlung in a strong radiation field. *Zhurnal Éksp. Teor. Fiz.* **1965**, *49*, 1215–1221. [*Sov. Phys. JETP* **1966**, *22*, 844–847].

[20] Kroll, N.M.; Watson, K.M. Charged-particle scattering in the presence of a strong electromagnetic wave. *Phys. Rev. A* **1973**, *8*, 804–809.

[21] Weingarsthofer, A.; Holmes, J.K.; Caudle, G.; Clarke, E.M. Direct observation of multiphoton processes in laser-induced free-free transitions. *Phys. Rev. Lett.* **1977**, *39*, 269–270.

[22] Wallbank, B.; Holmes, J.K. Laser-assisted electron-atom collisions. *Phys. Rev. A* **1993**, *48*, R2515–R2518.

[23] Mason, N.J. Laser-assisted electron atom collisions. *Rep. Prog. Phys.* **1993**, *56*, 1275–1346.

[24] Varró, S.; Ehlotzky, F. Remark on polarization effects in small-angle electron scattering by helium atoms in a $CO_2$ laser field. *Phys. Lett. A* **1995**, *203*, 203–208.

[25] Kanya, R.; Morimoto, Y.; Yamanouchi, K. Observation of laser-assisted electron-atom scattering in femtosecond intense laser field. *Phys. Rev. Lett.* **2010**, *105*, 123202.

[26] Morimoto, Y.; Kanya, R.; Yamanouchi, K. Laser-assisted electron diffraction for femtosecond molecular imaging. *J. Chem. Phys.* **2014**, *140*, 064201.

[27] Morimoto, Y.; Kanya, R.; Yamanouchi, K. Light-Dressing Effect in Laser-Assisted Elastic Electron Scattering by Xe. *Phys. Rev. Lett.* **2015**, *115*, 123201.

[28] McPherson, A.; Gibson, G.; Jara, H.; Johann, U.; Luk, T.; Mcintyre, I.; Boyer, K.; Rhodes, C. Studies of multiphoton production of vacuum-ultraviolet radiation in the rare gases. *J. Opt. Soc. Am. B* **1987**, *4*, 595–601.

[29] Ferray, M.; L'Huillier, A.; Li, X.F.; Lompre, L.A.; Mainfray, G.; Manus, C. Multiple-harmonic conversion of 1064 nm radiation in rare gases. *J. Phys. B At. Mol. Opt. Phys.* **1988**, *21*, L31.

[30] L'Huillier, A.; Balcou, P. High-order harmonic generation in rare gases with a 1-ps 1053-nm laser. *Phys. Rev. Lett.* **1993**, *70*, 774–777.

[31] Potvliege, R.M.; Shakeshaft, R. Multiphoton processes in an intense laser field: Harmonic generation and total ionization rates for atomic hydrogen. *Phys. Rev. A* **1989**, *40*, 3061–3079.

[32] Becker, W.; Long, S.; McIver, J.K. Higher-harmonic production in a model atom with short-range potential. *Phys. Rev. A* **1990**, *41*, 4112–4115.

[33] Ehlotzky, F. Harmonic generation in Keldysh-type Models. *Il Nuovo Cim.* **1992,** *14D*, 517–525.

[34] L'Huillier, A.; Balcou, P.; Candel, S.; Schafer, K.J.; Kulander, K.C. Calculation of high-order harmonic-generation processes in xenon at 1064 nm. *Phys. Rev. A* **1992**, *46*, 2778–2790.

[35] Corkum, P.B. Plasma perspective on strong field multiphoton ionization. *Phys. Rev. Lett.* **1993**, *71*, 1994–1997.

[36] Varró, S.; Ehlotzky, F. A new integral equation for treating high-intensity multiphoton processes. *Il Nuovo Cim.* **1993**, *15D*, 1371–1396.

[37] Lewenstein, M.; Balcou, P.; Ivanov, M.Y.; L'Huillier, A.; Corkum, P.B. Theory of high-harmonic generation by low-frequency laser fields. *Phys. Rev. A* **1994**, *49*, 2117–2132.

[38] Kuchiev, M.Y.; Ostrovsky, V.N. Quantum theory of high harmonic generation as a three-step process. *Phys. Rev. A* **1999**, *60*, 3111–3124.

[39] Farkas, G.; Tóth, C. Proposal for attosecond light pulse generation using laser induced multiple-harmonic conversion processes in rare gases. *Phys. Lett. A* **1992**, *168*, 447–450.

[40] Papadogiannis, N.A.; Witzel, B.; Kalpouzos, C.; Charalambidis, D. Observation of attosecond light localization in higher order harmonic generation. *Phys. Rev. Lett.* **1999**, *83*, 4289–4292.

[41] Paul, P.M.; Toma, E.S.; Breger, P.; Mullot, G.; Auge, F.; Balcou, P.; Muller, H.G.; Agostini, P. Observation of a train of attosecond pulses from high harmonic generation. *Science* **2001**, *292*, 1689–1692.

[42] Tzallas, P.; Charalambidis, D.; Papadogiannis, N.A.; Witte, K.; Tsakiris, G.D. Direct observation of attosecond light bunching. *Nature* **2003**, *426*, 267–271.

[43] Agostini, P.; DiMauro, L.F. The physics of attosecond light pulses. *Rep. Prog. Phys.* **2004**, *67*, 813–855.

[44] López-Martens, R.; Varjú, K.; Johnsson, P.; Mauritsson, J.; Mairesse, Y.; Saliéres, P.; Gaarde, M.B.; Schafer, K.J.; Persson, A.; Svanberg, S.; et al. Amplitude and phase cotrol of attosecond light pulses. *Phys. Rev. Lett.* **2005**, *94*, 033001.

[45] Sansone, G.; Benedetti, E.; Calegary, F.; Vozzi, C.; Avaldi, L.; Flammini, L.; Poletto, L.; Villoresi, P.; Altucci, C.; Velotta, R.; et al. Isolated single-cycle attosecond pulses. *Science* **2006**, *314*, 443–446.

[46] Krausz, F.; Ivanov, M. Attosecond physics. *Rev. Mod. Phys.* **2009**, *81*, 163–234.

[47] Wirth, A.; Hassan, M.T.; Grguraš, I.; Gagnon, J.A.; Moulet, A.; Luu, T.T.; Pabst, S.; Santra, R.; Alahmed, Z.A.; Azzeer, A.M.; et al. Synthesized light transients. *Science* **2011**, *334*, 195–200.

[48] Luu, T.T.; Garg1, M.; Kruchinin, S.Y.; Moulet, A.M.T.; Hassan, M.T.; Goulielmakis, E. Extreme ultraviolet high-harmonic spectroscopy of solids. *Nat. Lett.* **2015**, *521*, 498–502.

[49] Kühn, S.; Dumergue, M.; Kahaly, S.; Mondal, S.; Füle, M.; Csizmadia, T.; Farkas, B.; Major, B.; Várallyay, Z.; Cormier, E.; et al. The ELI-ALPS facility: The next generation of attosecond sources. *J. Phys. B At. Mol. Opt. Phys.* **2017**, *50*, 132002.

[50] Charalambidis, D.; Chikán, V.; Cormier, E.; Dombi, P.; Fülöp, J.A.; Janáky, C.; Kahaly, S.; Kalashnikov, M.; Kamperidis, C.; Kühn, S.; et al. The Extreme Light Infrastructure—Attosecond Light Pulse Source (ELI-ALPS) project. In *Progress in Ultrafast Intense Laser Science XIII*; Yamanouchi,

H.W.T., Paulus, G.G., Eds.; Springer International Publishing AG: Basel, Switzerland, 2017; pp. 181–218.

[51] Amini, K.; Biegert, J.; Calegari, F.; Chacón, A.; Ciappina, M.F.; Dauphin, A.; Efimov, D.K.; de Morisson Faria, C.F.; Giergiel, K.; Gniewek, P.; et al. Symphony on strong field approximation. *arXiv* **2018**, arXiv:1812.11447v1.

[52] Miloševic, D.B. Semiclassical approximation for strong-laser-field processes. *Phys. Rev. A* **2017**, *96*, 023413.

[53] Wikmark, H.; Guo, C.; Vogelsang, J.; Smorenburg, P.W.; Coudert-Alteirac, H.; Lahl, J.; Peschel, J.; Rudawski, P.; Dacasa, H.; Carlström, S.; et al. Spatiotemporal coupling of attosecond pulses. *Proc. Nat. Acad. Sci. USA* **2019**, *116*, 4779–4787.

[54] Nayak, A.; Dumergue, M.; Kühn, S.; Mondal, S.; Csizmadia, T.; Harshitha, N.G.; Füle, M.; Upadhyay-Kahaly, M.; Farkas, B.; Major, B.; et al. Saddle point approaches in strong field physics and generation of attosecond pulses. *Phys. Rep.* **2019**, *833*, 1–52.

[55] Van Vleck, J.H. The correspondence principle in the statistical interpretation of quantum mechanics. *Proc. Nat. Acad. Sci. USA* **1928**, *14*, 178–188.

[56] Feynman, R.P.; Hibbs, A.R. *Quantum Mechanics and Path Integrals*; McGraw-Hill: New York, NY, USA, 1965.

[57] Jain, M.; Tzoar, N. Compton scattering in the presence of coherent radiation. *Phys. Rev. A* **1978**, *18*, 538–545.

[58] Pert, G.J. The behaviour of atomic bound states in very strong electromagnetic fields. *J. Phys. B At. Mol. Phys.* **1975**, *8*, L173–L178.

[59] Basile, S.; Trombetta, F.; Ferrante, G. Twofold symmetric angular distribution in multiphoton ionization with elliptically polarized light. *Phys. Rev. Lett.* **1988**, *61*, 2435–2437.

[60] Basile, S.; Trombetta, F.; Ferrante, G.; Burlon, R.; Leone, C. Multiphoton ionization of hydrogen by a strong multimode field. *Phys. Rev. A* **1988**, *37*, 1050–1052.

[61] Ehlotzky, F. Remarks on Coulomb correction in scattering and ionization in a laser field. *Opt. Commun.* **1990**, *77*, 309–311.

[62] Kramers, H.A. *Collected Scientific Papers*; North-Holland Publishing Company: Amsterdam, The Netherland, 1956; p. 262.

[63] Henneberger, W.C. Perturbation method for atoms in intense light beams. *Phys. Rev. Lett.* **1968**, *21*, 838–841.

[64] Faisal, F.H.M. Collision of electrons with laser photons in a background potential. *J. Phys. B At. Mol. Phys.* **1973**, *6*, L312–L315.

[65] Gavrila, M. (Ed.) *Atoms in Intense Laser Fields*; Academic Press, Inc.: San Diego, CA, USA, 1992.

[66] Mittleman, M.H. *Introduction to the Theory of Laser-Atom Interactions*; Plenum Press: New York, NY, USA, 1993; Chapter 4.

[67] Reed, V.C.; Burnett, K.; Knight, P.L. Harmonic generation in the Kramers-Henneberger stabilization regime. *Phys. Rev. A* **1993**, *47*, R34–R37.

[68] Becker, A.; Faisal, F.H.M. Intense-field many-body S-matrix theory. *J. Phys. B At. Mol. Opt. Phys.* **2005**, *38*, R1–R56.

[69] Faisal, F.H.M. Strong-field S-matrix series with Coulomb wave final state. In *Progress in Ultrafast Intense Laser Science XIII*; Yamanouchi, H.W.T., Paulus, G.G., Eds.; Springer International Publishing AG: Basel, Switzerland, 2017; pp. 1–13.

[70] Bloch, F.; Nordsieck, A. Notes on the radiation field of the electron. *Phys. Rev.* **1937**, *52*, 54–59.

[71] Glauber, R.J. Some notes on multiple-boson processes. *Phys. Rev.* **1951**, *84*, 395–400.

[72] Schwinger, J. Theory of quantized fields. III. *Phys. Rev.* **1953**, *91*, 728–740.

[73] Varró, S. Theoretical Study of the Interaction of Free Electrons with Intense Light. Ph.D.Thesis, University of Szeged, Szeged, Hungary, 1981. *Hung. Phys. J.* **1983**, *31*, 399–454. (In Hungarian).

[74] Varró, S. Intensity effects and absolute phase effects in nonlinear laser-matter interactions. In *Laser Pulse Phenomena and Applications*; Duarte, F.J., Ed.; InTech: Rijeka, Croatia, 2010; Chapter 12, pp. 243–266.

[75] Antoine, P.; L'Hulillier, A.; Lewenstein, M. Attosecond pulse trains using high-order harmonics. *Phys. Rev. Lett.* **1996**, *77*, 1234–1237.

[76] Varró, S.; Farkas, G. Attosecond electron pulses from interference of above-threshold de Broglie waves. *Laser Part. Beams* **2008**, *26*, 9–19.

[77] Varró, S. Entangled photon-electron states and the number-phase minimum uncertainty states of the photon field. *New J. Phys.* **2008**, *10*, 053028.

[78] Varró, S. Entangled states and entropy remnants of a photon-electron system. *Phys. Scr.* **2010**, *T140*, 014038.

[79] Varró, S. Symmetric entangled coherent states yield ideal quantum attosecond pulses. In Proceedings of the 27th International Laser Physics Workshop (LPHYS'18), Nottingham, UK, 16–20 July 2018. Unpublished work, 2018.

[80] Gombkötő, A.; Czirják, A.; Varró, S.; Földi, P. Quantum-optical model for the dynamics of high-order-harmonic generation. *Phys. Rev. A* **2016**, *94*, 013853.

[81] Varró, S. A quantum concept of attosecond radiation: The attoquant. In Proceedings of the 7th International Conference on Attosecond Science and Technology (ATTO2019), Szeged, Hungary, 1–5 July 2019. Unpublished work, 2019.

[82] Gombkötő, A.; Varró, S.; Mati, P.; Földi, P. High-order harmonic generation as induced by a quantized field: Phase-space picture. *Phys. Rev. A* **2020**, *101*, 013418.

[83] Gorlach, A.; Neufeld, O.; Rivera, N.; Cohen, O.; Kaminer, I. The quantum-optical nature of high harmonic generation. *Nat. Commun.* **2020**, *11*, 4598.

[84] Gonoskov, I.; Tsatrafyllis, N.; Kominis, I.; Tzallas, P. Quantum optical signatures in strong-field laser physics: Infrared photon counting in high-order-harmonic generation. *Sci. Rep.* **2016**, *6*, 32821.

[85] Tsatrafyllis, N.; Kominis, I.K.; Gonoskov, I.A.; Tzallas, P. High-order harmonics measured by the photon statistics of the infrared driving-field exiting the atomic medium. *Nat. Commun.* **2017**, *8*, 15170.

[86] Tsatrafyllis, N.; Kühn, S.; Dumergue, M.; Földi, P.; Kahaly, S.; Cormier, E.; Gonoskov, I.; Kiss, B.; Varjú, K.; Varró, S.; et al. Quantum Optical Signatures in a Strong Laser Pulse after Interaction with Semiconductors. *Phys. Rev. Lett.* **2019**, *122*, 193602.

[87] Theocharis, L.; Lopez-Martens, R.; Haessler, S.; Liontos, I.; Kahaly, S.; Rivera-Dean, J.; Stammer, P.; Pisanty, E.; Ciappina, M.F.; Lewenstein, M.; et al. Quantum-optical spectrometry in relativistic laser–plasma interactions using the high-harmonic generation process: A proposal. *Photonics* **2021**, *8*, 192.

[88] Burenkov, I.A.; Tikhonova, O.V. Features of multiphoton-stimulated bremsstrahlung in a quantized field. *J. Phys. B At. Mol. Opt. Phys.* **2010**, *43*, 235401.

[89] Bogatskaya, A.; Volkova, E.; Popov, A. Spontaneous transitions in atomic system in the presence of high-intensity laser field. *EPLA* **2016**, *116*, 14003.

[90] Bogatskaya, A.; Volkova, E.; Popov, A. Spectroscopy of the atomic system driven by high intensity laser field. *arXiv* **2017**, arXiv:1701.05777v1.

[91] Bogatskaya, A.V.; Volkova, E.A.; Popov, A.M. Spontaneous emission ofatoms in a strong laser field. *JETP* **2017**, *125*, 587–596.

[92] Akhiezer, A.I.; Berestetskii, V.B. *Quantum Electrodynamics*; Interscience Publisher: New York, NY, USA, 1965.

[93] Białynicki-Birula, I.; Białynicki-Birula, Z. *Quantum Electrodynamics*; Pergamon Press: Oxford, UK.; Warszawa, Poland, 1975.

[94] Loudon, R. *The Quantum Theory of Light*; Clarendon Press: Oxford, UK, 2000.

[95] Schleich, W.P. *Quantum Optics in Phase Space*; Wiley-VCH: Weinheim, Germany, 2001.

[96] Scully, M.O.; Zubairy, M.S. *Quantum Optics*; Cambridge University Press: Cambridge, UK, 1997.

[97] Loudon, R.; Knight, P.L. Squeezed light. *J. Mod. Opt.* **1987**, *34*, 709–759.

[98] Born, M.; Wolf, E. *The Principles of Optics*; Cambridge University Press: Cambridge, UK, 2009; 7th, Expanded, Fifth Printing.

[99] Cooper, J.W. Photoionization from outer atomic subshells. A model study. *Phys. Rev*. **1962**, *128*, 681–693.

[100] Shiner, A.D.; Schmidt, B.E.; Trallero-Herrero, C.; Corkum, P.B.; Kieffer, J.-C.; Légaré, F.; Villeneuve, D.M. Observation of Cooper minimum in krypton using high harmonic spectroscopy. *J. Phys. B At. Mol. Opt. Phys*. **2012**, *45*, 074010.

[101] Shiner, A.D.; Schmidt, B.E.; Trallero-Herrero, C.; Wörner, H.-J.; Patchkovskii, S.; Corkum, P.B.; Kieffer, J.-C.; Légaré, F.; Villeneuve, D.M. Probing collective multi-electron dynamics in xenon with high-harmonic spectroscopy. *Nat. Phys.* **2011**, *7*, 464–467.

[102] Higuet, J.; Ruf, H.; Thiré, N.; Constant, R.; Cormier, E.; Descamps, D.; Mével, E.; Petit, S.; Pons, B.; Mairesse, Y.; et al. High-harmonic spectroscopy of the cooper minimum in argon: Experimental and theoretical sudy. *Phys. Rev. A* **2011**, *83*, 053401.

[103] Schoun, S.B.; Chirla, R.; Wheeler, J.; Roedig, C.; Agostini, P.; DiMauro, L.F. Attosecond pulse shaping around a Cooper minimum. *Phys. Rev. Lett.* **2014**, *112*, 153001.

[104] Glauber, R.J. The quantum theory of optical coherence. *Phys. Rev*. **1963**, *130*, 2529–2539.

[105] Glauber, R.J. Coherent and incoherent states of the radiation field. *Phys. Rev*. **1963**, *131*, 2766–2788.

[106] Schrödinger, E. Der stetige Übergang von der Mikro-zur Makromechanik. *Die Nat.* **1926**, *14*, 664–666.

[107] Dodonov, V.V., Man’ko, V.I., Eds.; *Theory of Nonclassical States of Light,* Taylor & Francis: London, UK; New York, NY, USA, 2003)

[108] Andersen, U.L.; Gehring, T.; Marquardt, C.; Leuchs, G. 30 years of squeezed light generation. *Phys. Scr*. **2016**, *91*, 053001.

[109] Neumann, J. *Mathematische Grundlagen der Quantenmechanik*; Springer: Berlin, Germany, 1932.

[110] Bargmann, Y.V.; Butera, P.; Girardello, L.; Klauder, J.R. On the completeness of the coherent states. *Rep. Math. Phys.* **1971**, *2*, 221–228.

[111] Perelomov, A.M. Remark on the completeness of the system of coherent states (in Russian). *Teor. Mat. Fiz*. **1971**, *6*, 213–224.

[112] Gradshteyn, I.S.; Ryzhik, I.M. *Table of Integrals, Series and Products*, 6th ed.; Academic Press: San Diego, CA, USA, 2000.

[113] Lewenstein, M.; Ciappina, M.F.; Pisanty, E.; Rivera-Dean, J.; Lamprou, T.; Tzallas, P. The quantum nature of light in high harmonic generation. *arXiv* **2020**, arXiv:physics.optics/2008.10221.

[114] Ozawa, A.; Davila-Rodriguez, J.; Bounds, J. R.; Schuessler, H. A.; Hänsch, T. W.; Udem, Th. Single ion fluorescence excited with a single mode of an UV frequency comb. *Nature Communications* **2017**, *8*, 44.

[115] Wood, W.M.; Siders, C.W.; Downer, M.C. Measurement of femtosecond ionization dynamics of atmospheric density gases by spectral blueshifting. *Phys. Rev. Lett.* **1991**, *67*, 3523–3526.

[116] Rae, S.C.; Burnett, K. Detailed simulation of plasma-induced spectral blueshifting. *Phys. Rev. A* **1992**, *46*, 1084–1090.

[117] Orfanos, I.; Skantzakis, E.; Liontos, I.; Tzallas, P.; Charalambidis, D. Ponderomotive shifts by intense laser-driven coherent extreme ultraviolet radiation. *J. Phys. B At. Mol. Opt. Phys.* **2021**, *54*, 084002.

[118] Dattoli, G.; Del Franco, M.; Labat, M.; Ottaviani, P.L.; Pagnutti, S. Introduction to the physics of free electron laser and comparison with conventional sources. In *Free Electron Lasers*; Varró, S., Ed.; InTech: Rijeka, Croatia, 2012; Chapter 1, pp. 1–38.

[119] Dattoli, G.; Nguyen, F. Free Electron Laser and Fundamental Physics. *Prog. Part. Nucl. Phys.* **2018**, *99*, 1–28.

[120] Debus, A.; Steininger, K.; Kling, P.; Carmesin, C.M.; Sauerbrey, R. Realizing quantum free-electron lasers: A critical analysis of the experimental challenges and theoretical limits. *Phys. Scr.* **2019**, *94*, 074001.